\documentclass[11pt,a4paper]{article}
\usepackage{jheppub}
\usepackage{color}
\usepackage{graphics}
\usepackage{amsfonts,amsbsy,latexsym,amssymb,amscd,amstext}
\usepackage{epsfig}
\usepackage{graphicx}
\usepackage{subcaption}
\usepackage{subfloat}
\makeatletter
\def\@fpheader{\relax}
\makeatother
\newcommand{\Tr}{\mathrm{Tr}}
\newcommand{\E}{E}
\newcommand{\be}{\begin{equation}}
\newcommand{\ee}{\end{equation}}
\newcommand{\bea}{\begin{eqnarray}}
\newcommand{\eea}{\end{eqnarray}}

\title{Testing volume independence of SU(N) pure gauge theories at
large N}
\author{Antonio  Gonz\'alez-Arroyo $^{a-b}$ and Masanori Okawa $^{c}$ \\
  $^a$ Instituto de F\'{\i}sica Te\'orica UAM/CSIC \\
   Nicol\'as Cabrera 13-15 \\
                 Universidad Aut\'onoma de Madrid, \\
		 Cantoblanco E-28049--Madrid, Spain \\
 $^b$	Departamento de F\'{\i}sica Te\'orica, C-15 \\
                  Universidad Aut\'onoma de Madrid, \\
		                   Cantoblanco E-28049--Madrid, Spain \\
		 $^c$ Graduate School of Science \\
		 Hiroshima University \\
		 Higashi-Hiroshima, Hiroshima 739-8526, Japan \\
		 
				         }
						    
\abstract{
In this paper we present our  results concerning the dependence of 
Wilson loop expectation values on the size of the lattice and the rank 
of the SU(N) gauge group. This allows to test the claims about  volume
independence in the large N limit, and the crucial dependence on
boundary conditions. Our highly precise results provide 
strong support for the validity of the twisted reduction mechanism 
and the TEK model, provided the fluxes are chosen within the appropriate
domain. 
}

	     
	                  \preprint{IFT-UAM/CSIC-14-102; FTUAM-14-39;
			  HUPD-1406 }

\begin{document}
\maketitle

\section{Introduction}
Gauge theories lie at the basis of our understanding of Particle
Physics. They are also  beautiful mathematical models which are full 
of rich phenomena, both perturbative and non-perturbative. The pure
gluon  version of the theory has no free parameters and its dynamics 
underlies many of the fascinating properties of the interaction of
quarks, such as Confinement and chiral  symmetry breaking. 
Despite having no scale in its formulation, the theory is not conformal 
invariant because a scale $\Lambda$ is generated by the mechanism of 
dimensional transmutation~\cite{CW}. All physical properties 
of the theory can then be expressed, depending on its dimensions, in 
the appropriate units given by powers of $\Lambda$. Observable quantities
which involve large energy scales (with respect to $\Lambda$) can be 
calculated by perturbative methods, thanks to the property of
asymptotic freedom~\cite{politzer}-\cite{GW}. On the contrary, if some low
energy phenomena is involved, perturbation theory fails dramatically 
and one has to use non-perturbative techniques. Fortunately a 
very powerful formulation was introduced by K. Wilson~\cite{wilson},
allowing the development of several non-perturbative methods of
computation. Lattice gauge theories define the quantum field theory as
the limit of an statistical mechanical system at critically. One 
can then employ numerical methods which were originally devised to
handle statistical mechanical systems. Unfortunately, for the study of
the critical behaviour only numerical Monte Carlo methods seem useful. 
Nonetheless, progress in computer technology and in numerical
algorithms have made possible the calculation of several observables 
of Yang-Mills and the more demanding theories with dynamical fermions
(for recent results see Ref.~\cite{latticeconf}). 

In the 70's G. `t Hooft~\cite{LN} suggested the use of a new expansion
parameter: $1/N$, the inverse of the rank of the group. 
Asymptotic freedom, Confinement, dimensional transmutation and  a massive 
spectrum are still present  at the lowest order approximation: the large N limit. 
Furthermore, it is known that several aspects of the theory are
simpler in that limit: planar graphs, stable spectrum, 
factorisation~\cite{migdal}, etc. It seems  clear that a more profound 
understanding of the large N limit would precede that of finite N. 
The large N limit also plays a crucial simplificatory role in the new
ideas and methods that have emerged under the heading of 
AdS/CFT~\cite{ads}.

One of the obvious questions is then whether the large N limit also 
introduces  a simplification in combination with the lattice gauge
theory formulation. At first, it seems just the opposite. Numerical 
methods are devised to work with finite N values.  Results at the
large N limit  then follow   from extrapolating those obtained at  
finite $N$. But, the  numerical  effort grows with the number of 
degrees of freedom (hence, at least,  as $N^2$). Nevertheless, following 
this procedure might be worth  the effort, in order  to produce 
predictions that can be matched with  those of other approaches. 
A large literature has been generated following this line which can be 
consulted in Ref.~\cite{LP}. 

There is, however, one potentially 
simplificatory phenomenon which was discovered when combining 
large $N$ with the lattice approach: {\em reduction} or {\em volume
independence}.  The idea emerged from a crucial observation made by
Eguchi and Kawai(EK)~\cite{EK} when looking at the loop equations~\cite{migdalmakeenko}
obeyed by Wilson loops on the lattice. Assuming invariance under 
$Z^4(N)$ centre symmetry and factorisation, they concluded that these 
equations are independent of the lattice volume. 

Thus, if the phenomenon is correct, one can reduce the study of the
large N limit to a simple matrix model with only $d$ matrices ($d$ 
is the space-time dimension). This is, no doubt, a very powerful 
simplification, at least at the conceptual level. The first question is then 
whether the idea is indeed correct. However, there are other secondary 
questions which could be crucial for applications. In particular, one
might inquire about the size and nature of the corrections. What are the
leading corrections that  one has at large but finite N? 
These are  the central issues which our paper tries to address. 

The history of the subject is long. The initial stages took place soon
after the EK paper. It was shown that the EK model, obtained by reducing
to a one-point  lattice with periodic boundary conditions,
violates the centre-symmetry  condition for reduction at weak 
coupling~\cite{bhanot}. 
Indeed, the one point lattice had been studied before the EK paper in 
Ref.~\cite{gajka}, and the weak coupling path  integral was shown to be 
dominated by  symmetry breaking configurations. 
Fortunately, the authors of Ref.~\cite{bhanot} proposed a modification 
under the name Quenched Eguchi-Kawai model (QEK) devised to restore
the symmetry and, hence, the validity of reduction. On the other hand, 
the studies of Ref.~\cite{gajka} suggested that boundary conditions have a
very strong influence on the weak coupling behaviour at finite volume. 
Thus, the present authors\cite{TEK1}-\cite{TEK2} presented a new version
of reduction based upon twisted boundary conditions and aimed  at
respecting a sufficiently large subgroup of the centre symmetry group. 
The new model, called Twisted Eguchi-Kawai model
 (TEK), follows by imposing twisted boundary conditions and subsequently
reducing the model to one point. At large volumes the boundary
conditions do not influence the dynamics of the model, but at small
volumes they are crucial. 

Both the QEK and the TEK model passed all tests of absence of centre
symmetry breaking done at the time of their proposal, and looked 
as valid implementations of the reduction idea. Unfortunately, 
attempts to solve these models explicitly failed. With an increasing 
interest in large N gauge theories the situation was re-examined in
recent times, and signals of symmetry breaking were observed in both
cases. By construction the QEK model is invariant under a certain
subgroup of the $Z^4(N)$ group. However, there are possible subgroups 
associated with the action  on several directions simultaneously 
that can be violated and were observed to do so in Ref.~\cite{sharpe}.
In the case of the TEK model, although the remnant symmetry 
cannot be broken at infinite weak coupling, it can still be broken at 
intermediate values. The first observations of breaking were obtained by
one of the present authors and collaborator~\cite{TIMO}. Later 
on, several other authors~\cite{TV}-\cite{Az}
confirmed the breaking and showed how this problem affects crucially the
possibility of reduction in the continuum limit.  

During these years there have been other attempts to recover the
validity of the reduction idea and its simplificatory character. 
Some authors argued that other large N gauge theories 
might be free from the symmetry breaking problems that invalidated
the reduction of the EK model. In particular, theories with several 
flavours of fermions in the adjoint representation~\cite{KUY} are good candidates.
This has triggered several efforts to obtain information about 
these theoretically interesting theories, by benefiting from the 
computational advantage of their reduced  
versions~\cite{AHUY,BS2,BKS,HN1,HN2,AGAMOAdj1,AGAMOAdj2,AGAMOAdj3}. 
Although this is certainly very interesting, here we will not consider  
these theories any further since our interest in this paper is centred 
upon pure gauge theories.

Another approach was proposed by Narayanan and Neuberger~\cite{NN}, 
lately referred as {\em partial reduction}. The idea 
is to combine space and group degrees of freedom in an efficient way. 
Their study shows that at any value of the coupling there is a range 
of values of  $L$ ($L$ being the linear size of the lattice box) 
for  which the symmetry is respected at large $N$. More precisely, there 
is a minimum  box size $L_c(b)$, which grows as the coupling gets weaker. 
Infinite volume large $N$ results can be approached by tuning $L$ and $N$
appropriately. Although this  idea was mostly developed  with 
periodic boundary conditions, it is  also possible to  combine it  with
twisted boundary conditions~\cite{GANN}. This is a possibility that we 
will analyse further in the present paper. 

Reasons for the failure of the TEK model were given in
Ref.~\cite{TIMO}-\cite{TV}. Starting from these studies, the present authors 
proposed a simple modification of the original procedure which 
could restore the symmetry and recover the validity of
reduction~\cite{TEK3}. The idea is to tune the fluxes appropriately 
when taking the large $N$ limit. Indeed, the twisted boundary
conditions are not unique; they depend upon the choice of 
discrete fluxes	  through each plane. In Ref.~\cite{TEK2} we determined the
range of fluxes that guarantee the preservation of the symmetry at
very weak coupling. However, it is now clear that the actual choice of fluxes
within this range has an influence at intermediate values of the coupling.
In Ref.~\cite{TEK3} we argued that, with a suitable choice, the problems 
can be avoided at all  values of the coupling. 

In the last few years we have set up a program to study this problem. 
As explained earlier, the question is two-fold: Is reduction correct?
Is it useful? The second aspect might be crucial for its ultimate
interest as a valid simplification. Trying to answer the second
question directly, we set ourselves the goal of computing an observable of the
large $N$ continuum infinite volume theory using the TEK model. For 
simplicity we  focused on the behaviour of large Wilson loops and 
the string tension.  In parallel with the use of the reduced model,
we also used the traditional extrapolation methodology to obtain the
same quantity.  Our comparison was presented in
Ref.~\cite{TEK4}-\cite{TEK5},
and showed that the  extrapolation of the finite $N$ string tension 
agreed  within the
percent level with the quantity obtained from the reduced model. 
The result also matched reasonably well with other determinations of the
large N string tension~\cite{ATT}-\cite{lohmayer}.

Obtention of the continuum string tension is an elaborate procedure 
involving noise reduction techniques on the raw data, extrapolation to 
large size Wilson loops and then   to the continuum limit. All of
these steps, as well as the traditional methodology, involve
extrapolations. This is always a dangerous arena, although it 
could hardly be a coincidence  that the results match so well. 
Our initial focus on  continuum results  is understandable, since after all
the lattice is primarily  a method to obtain properties of the continuum 
formulated Yang-Mills theory. However, the reduction idea should be  also 
valid for the lattice  model itself, even  away from criticality. In this
paper we want precisely to 
look into that. We will focus  upon simple observables of the lattice
theory. Some are robust, highly precise, and devoid of any technical
issues which might shade the validity of the result. Thus, rather than 
testing reduction by checking the preservation of the symmetry needed to
validate the original proof, we will produce  a direct test by
comparing the volume independence of the corresponding observables.
Furthermore, this is a quantitative comparison which can put limits on
the errors committed. Ultimately, these errors are crucial for a precision
determination of observables. 

What do we know about the finite $N$ corrections to reduction?
Perturbation theory gives some information in this respect. The
propagators of the TEK model turn out to be identical to those of
a lattice of size $(\sqrt{N})^4$~\cite{TEK2}. This is reasonable since
we are essentially mimicking  the  space-time degrees of freedom by those
of the group. This seems more efficient than the equivalent  counting for
the QEK model, which suggests an effective volume of $(N^{1/4})^4$.
Furthermore, if we opt for partial reduction in the twisted case, the 
effective length of the box is given by $L\sqrt{N}$. In addition to
the effect on the propagators,  the Feynman rules for the vertices also
adopt a peculiar form. The structure constants turn into complex phases 
which depend on the ``effective momentum'' degrees of freedom. It is
precisely this structure which suppresses the non-planar diagrams of
the theory. The choice of flux enters the explicit form of the phase
factors through a particular rational number in the exponent. Since
all the factors cancel out for planar diagrams we were guided into
thinking that the choice of fluxes was irrelevant. Now we know that
this choice might  be crucial in guaranteeing the preservation of centre symmetry. 
Recently~\cite{GPGAO1,GPGAO2,GPGAO3,GPGAO4} we examined the problem with 
more detail for the simpler case of the 2+1 dimensional gauge theory. Physical
observables seem to depend smoothly on the rational factor. This
suggests a stronger  form of volume independence valid at finite $N$. 
According to the previous considerations, as long as we preserve the 
effective spatial size ($LN$ in 2 dimensions) and the rational factor 
in the exponent of the vertices,  we can trade spatial degrees of 
freedom by group degrees of freedom after a suitable tuning of the fluxes. It would be
interesting to examine the situation for the 3+1 dimensional theory.

An important new ingredient appeared on the scene after the scientific
community generated interest into quantum field theories in
non-commutative space-times (for a review see Ref.~\cite{nekrasov}).
It turns out that the afore-mentioned Feynman rules of reduced models
are a discretized version of those appearing in those type of theories, 
where the rational number mentioned earlier is just proportional to
the non-commutativity parameter. Indeed, the continuous version of the
rules were actually anticipated by a generalisation of the reduced model 
obtained by one of us~\cite{AGAKA}, and which captures the
non-commutativity in its formulation. All these new connections suggest
that  the results obtained at finite $N$ might have an interpretation
and a usefulness in their own right.

In the previous paragraphs we have set up the situation that acts as
background of the present paper. In the next section we will present 
all the methodological details leading to our results. This includes 
the description of the models and the parameters involved in them, the 
observables that we are going to study and the details about the 
simulations. The following section contains the analysis of  our 
results for Wilson loops.  The paper ends with a short concluding
section. 

\section{Pure gauge theory on a finite lattice}
\subsection{The Models}
As mentioned in the introduction we will focus upon pure gauge 
lattice theory at finite volume. The dynamical variables of our 
theory are the link variables $U_\mu(n)$ which are elements of the 
SU(N) group. The index $\mu$ ranges over the 4 directions of space-time
$0,1,2,3$, while $n$ runs through all points of an $L^4$
hypercubic lattice $\mathcal{L}$. The partition function of the model is given by 
\be
Z= \prod_{n\in \mathcal{L}} \prod_{\mu=0}^3 \left( \int dU_\mu(n)\right)
e^{-S(U_\mu)}
\ee
where the link variables are integrated with the Haar measure of the
SU(N) group, and $S$ is action of the model. In this paper we will
choose the simplest version of the action, proposed by Wilson,
\be
\label{Wilson_action}
S_W(U_\mu) =-b N \sum_{n\in \mathcal{L}} \sum_{\mu\ne\nu} \Tr(U_{\mu
\nu}(n))
\ee
where $U_{\mu  \nu}(n)$ is the unitary matrix corresponding to an
elementary plaquette starting at point $n$ and living in the $\mu-\nu$
plane. Its expression in terms of the link variables is
\be
\label{Umunu}
U_{\mu  \nu}(n) = U_\mu(n) U_\nu(n+\hat{\mu})U^\dagger_\mu(n+\hat{\nu}) U^\dagger_\nu(n)
\ee
where $\hat{\mu}$ is the unit vector in the $\mu$ direction.
The lattice coupling $b$ is the lattice counterpart of the 
inverse of `t Hooft coupling $\lambda=g^2N$. Thus, the large $N$ limit 
has to be taken keeping $b$ fixed. In the previous formulas, we
assumed periodic boundary conditions for the variables:
$U_\mu(n+Lm)=U_\mu(n)$, for any integer vector $m$. Thus, the lattice
is rather a toroidal one. 

Wilson action is invariant under gauge transformations of the link variables:
\be
U_\mu(n) \longrightarrow \Omega(n) U_\mu(n) \Omega^\dagger(n+\hat{\mu})
\ee
where $\Omega(n)$ are arbitrary SU(N) matrices.
All physical observables are required to be gauge invariant. This property 
allows to relax the periodicity condition to 
\be
U_\mu(n+Lm)=\Omega(n,m)U_\mu(n)\Omega^\dagger(n+\hat{\mu},m)
\ee
To make this formula well-defined, the same gauge transformation has to
result from two different ways to reach the same replica  point. 
Since matrices do not necessarily  commute, this implies consistency 
conditions to be  satisfied by the gauge transformation matrices 
$\Omega(n,m)$. In particular, one can consider $n$ being the origin 
and $m=\hat{\mu}+\hat{\nu}$, 
and construct the gauge transformation in terms of the gauge
transformations for $m=\hat{\mu}$ and $m=\hat{\nu}$.
The consistency condition reads:
\be
\Omega(L\hat{\mu},\hat{\nu}) \Omega(0,\hat{\mu})= z_{\mu \nu}
\Omega(L\hat{\nu},\hat{\mu}) \Omega(0,\hat{\nu})
\ee
with $z_{\mu \nu}=\exp\{2\pi i n_{\mu \nu}/N\} $ an element of the
centre of the SU(N) group. Notice that if $n_{\mu \nu}=0$ for all
planes, one can choose all $\Omega(n,\hat{\mu})$ to be the identity and we recover
periodic boundary conditions. On the other hand, if some $n_{\mu \nu}\ne 0$
we get the so-called twisted boundary conditions introduced by `t
Hooft~\cite{TB}. The integer $n_{\mu \nu}=-n_{\nu \mu}$ can be interpreted 
as a  discrete flux modulo $N$ through the $\mu-\nu$ plane. 

Now one can perform a change of variables to new link variables which 
are periodic at the expense of introducing a phase factor at certain
plaquettes. The Wilson action   now adopts the form
\be
\label{Wilson_action_twist}
S_W(U_\mu) =-b N \sum_{n\in \mathcal{L}} \sum_{\mu\ne\nu} z_{\mu
\nu}(n)\Tr(U_{\mu  \nu}(n))
\ee
The $z_{\mu \nu}(n)$ factors can be moved around depending on the
change of variables, but their product
over each plane cannot be changed. One possibility is to make  all
factors equal to one except  for a single plaquette sitting at the corner
of each plane.  It is quite clear that for large volumes this has little
influence in
the local dynamics, but at small volumes it has an important effect. 

An extreme situation occurs in reduced models. The lattice  consists of 
a single point $L=1$. One can drop the $n$ dependence of links and
plaquettes and one is left with a  system involving only $d$ SU(N)
matrices $U_\mu$. If one takes periodic boundary conditions one has
the EK model. If, on the contrary, one chooses twisted boundary
conditions, one has the TEK model. The latter has more freedom because
one can take different values for the fluxes $n_{\mu \nu}$ having
different properties. For a class of values there exist configurations 
having zero-action. Thus, in these cases, there exist matrices
$U_\mu=\Gamma_\mu$ such that they satisfy 
\be
\Gamma_\mu \Gamma_\nu= z_{\nu \mu} \Gamma_\nu \Gamma_\mu
\ee
These are the twist-eaters~\cite{gjka}-\cite{AF}. We will also impose 
the condition that the $\Gamma_\mu$ matrices generate an irreducible
algebra. The interested reader can consult the literature to see the
conditions imposed on the values of $n_{\mu \nu}$ to achieve this goal
(see for example Ref.~\cite{AGA}). Rather than working with the
most general case we will concentrate on a rather simple situation
which tries to preserve as much symmetry as possible among the
different directions: the symmetric twist. The situation occurs
whenever $N$ is the square of an integer $N=\hat{L}^2$. Since, our goal 
is the large N limit, this restriction is not problematic. Furthermore, 
we will usually take $\hat{L}$ to be a prime number to avoid
$Z(\hat{L})$ from having proper subgroups. In summary, for the
symmetric twist case the twist factors for each plane are:
\be
z_{\mu \nu}= e^{2 \pi i k/\hat{L}} \quad \mathrm{ for }\  \mu<\nu 
\ee
where $\hat{L}=\sqrt{N}$ and $k$ is an integer defined modulo $\hat{L}$.
Choosing $k=0$ we recover the case of periodic boundary conditions.  

For completeness let us specify the form of the $\Gamma_\mu$ matrices 
for a symmetric twist with arbitrary $k\ne 0 $.
To do so, we first  introduce two $\hat{L} \times \hat{L}$ matrices
$P_{\hat{L}}$ and $Q_{\hat{L}}$ whose non-zero matrix elements are
given by
\bea
P_{\hat{L}}(\ell,\ell+1) = 1\ \ (\mathrm{for}\ 1\le \ell \le \hat{L}-1) ,\ \
P_{\hat{L}}(\hat{L},1)=1   
\\
 Q_{\hat{L}}(\ell,\ell) = \exp{\pi i k\left[ {1-\hat{L} \over
\hat{L}} + {2(\ell-1) \over \hat{L}}  \right] }
\ \ (\mathrm{for}\ 1\le \ell \le \hat{L}) 
\eea

\noindent
They are SU($\hat{L}$) matrices satisfying

\be
P_{\hat{L}} Q_{\hat{L}} = \exp{\left( {2\pi i k\over \hat{L}} \right)}
Q_{\hat{L}} P_{\hat{L}} 
\ee

\noindent
The four $N \times N$ matrices $\Gamma_{\mu}$ with $N=\hat{L}^2$,
are then given by the direct product of $P_{\hat{L}}$ and $Q_{\hat{L}}$ as
follows
\bea
\nonumber
\Gamma_{0} &=& Q_{\hat{L}} \otimes Q_{\hat{L}}\\
\nonumber
\Gamma_{1} &=& Q_{\hat{L}}\otimes P_{\hat{L}} Q_{\hat{L}} \\
\Gamma_{2} &=& Q_{\hat{L}} \otimes P_{\hat{L}} \\
\nonumber
\Gamma_{3} &=& P_{\hat{L}} \otimes I_{\hat{L}} 
\eea
\noindent
with $I_{\hat{L}}$ the $\hat{L} \times \hat{L}$ unit matrix.

\subsection{Observables}
Now let us consider the main observables. The most natural gauge
invariant observables are Wilson loops. A  path on the lattice 
is a finite sequence of oriented links, such that the endpoint of
one element in the sequence coincides with the origin of the next
element. For each  path we can construct a unitary matrix as a
product of the associated link variables following, left to right, the
order of the sequence. The reverse path is associated with the inverse
matrix. A closed path is a path such that the endpoint of the last
element in the sequence coincides with the origin of the first element. 
The trace of the corresponding unitary matrix is just the corresponding 
Wilson loop , which is gauge invariant. 

It is clear that the simplest non-trivial closed path is an elementary
plaquette $P_{\mu \nu}(n)$. The associated unitary matrix 
$U_{\mu \nu}(n)\equiv U(P_{\mu \nu}(n))$ has the expression given
earlier (Eq.~\ref{Umunu}). The corresponding expectation value 
$\E$ is the  most precise lattice observable
\be
\E= \frac{1}{N}  \langle \Tr(U_{\mu \nu}(n))\rangle
\ee
where the averaging is over positions,  orientations and
configurations.  In pure gauge theories with Wilson action, the
quantity is connected
with the mean value of the action as follows
\be
\E=\frac{1}{Vd(d-1)bN^2}\langle S_W\rangle
\ee
where $V$ is the lattice volume and $d$ the space-time dimension.

For the case of twisted boundary conditions the change of variables 
into periodic link variables transforms the plaquette observable 
as follows:
\be
U_{\mu \nu}(n) \longrightarrow z_{\mu \nu}(n) U_{\mu \nu}(n)
\ee
where the centre element $z_{\mu \nu}(n)$ is the same one appearing in
the action. The above replacement  in the formula for $\E$ gives the 
correct expression with twisted boundary conditions.

A priori,  $\E$ is a function of the different parameters entering the
simulation $\E(b,N,L,n_{\mu \nu})$. Given its local character this 
quantity is expected to have a well-defined infinite volume limit 
which is independent of the boundary conditions: $\E(b,N)$.
On general grounds one expects it to have also a well-defined large $N$
limit $\E_\infty(b)$, with corrections that go as powers of $1/N^2$.

Apart from the plaquette, one can consider other  Wilson loops. Here
we will focus on those associated to a rectangular path of size 
$R\times T$. The corresponding expectation value will be named
$W(R,T)$.
%\be
%W(R,T)=
%\ee
Once more, the quantity has to be modified in the standard variables 
for twisted boundary conditions. The modification is just to multiply
the expression by the product of all $z_{\mu \nu}(n)$ factors for 
the plaquettes contained in the rectangle. 

Just as for the plaquette, the expectation value will depend on 
$b$, $N$, $L$ and $n_{\mu \nu}$, but also depends on $R$ and $T$. One expects 
a well-defined thermodynamic limit as $L$ goes to $\infty$ at fixed
$R$ and $T$. For obvious reasons, the finite volume corrections should
be larger as $R$ and $T$ get closer to $L$. These corrections should 
depend on the boundary conditions. An extreme situation occurs for the
reduced model. In that case, the loop extends beyond the size of the box:
\be
W(R,T)= \frac{1}{N}z_{\mu \nu}^{RT} \langle \Tr(U_\mu^R
U_\nu^TU_\mu^{-R} U_\nu^{-T})\rangle
\ee
Nevertheless, the statement of volume independence at large $N$
implies that even loops that extend beyond the size of the box should 
behave as those obtained at infinite volume.

\subsection{Methodology}

The numerical results presented in this paper were obtained using
a heat-bath Monte Carlo simulation. For the Wilson
action case, each link variable $U_\mu(n)$ has a distribution determined by the
part of the action containing it, which has  the form 
\begin{equation}
\label{distrib}
 {\rm ReTr}\{U_\mu(n) V_\mu(n)\}
 \end{equation}
where $V_\mu(n)$ is constructed  in terms of the neighbouring links. 
Thus, at each local update one should generate a new link according to the 
previous distribution. However, the constraint that $U_\mu(n)$ belongs
to SU(N) complicates this task.   As has been demonstrated in \cite{MO}, 
a SU(N) link variable can be updated by successive multiplication of SU(2) 
sub-matrices. However, the standard Creutz' s heat
bath-algorithm has a very low acceptance rate for SU(N) gauge theories
with N larger than 8. To avoid  this, we have used a heat-bath
algorithm proposed by Fabricius and Haan \cite{FH} that significantly 
improves the acceptance rate for larger values of N.

Apart from this,  there are two technical points which significantly speed
up the SU(2) heat bath manipulations.
\begin{enumerate}
\item Suppose $A$ is an SU(2) sub-matrix, and we want to change the
link variable as follows $U_\mu(n)'= A U_\mu(n)$. 
During the updating step, we should also store $W=U_\mu(n) V_\mu(n)$,
namely $W'=AW$, which  only requires $O(N)$ arithmetic.
\item An  SU(N) matrix has $N(N-1)/2$ SU(2) sub-matrices (i,j).   For
odd(even) N,
there are $n_1=N$ ($N-1$) sets having $n_2=(N-1)/2$ ($N/2$)   (i,j) pairs,  which can be
manipulated simultaneously.  For example, for N=5, we have the
following 5 sets of two pairs: 
$\{(1,2),(3,4)\}$, $\{(1,3),(2,5)\}$, $\{(1,4),(3,5)\}$, 
$\{(1,5),(2,4)\}$, $\{(2,3),(4,5)\}$.
If the computer supports vector or multi-core coding,  we should make
full  use of these $n_2$ independent manipulations. 
\end{enumerate}

For the case of the reduced model (having $L=1$), the previous
methodology fails, because the distribution is not given by
Eq.~\ref{distrib}. The part of the action containing $U_\mu(n)$
depends quadratically rather than linearly on it. However, one
can return to a linear dependence by introducing a Gaussian random
matrix, which after integrating over it reproduces the original
form~\cite{FH}. Thus, we can alternate the updating of the new random
matrix with the previous heat-bath update.

Finally a link update is obtained by applying the SU(2) update for
each of the $N(N-1)/2$ SU(2) subgroups in SU(N). A sweep is obtained by 
applying an update for all the links on the lattice, followed by 5 
over-relaxation updates.

To reduce autocorrelations  we typically measured   configurations separated
by 100 sweeps. In any case, the final errors
were estimated by a jack-knife method with much larger separation
among groups. Another important point concerns thermalization.
Typically we discarded the initial 10\% of our 
configurations. We monitored several quantities to ensure that we
observed no transient effects in the resulting data. Related to this
point there is the choice  of initial configuration. For the TEK model 
we always initiated thermalization  from  the cold configuration that 
minimises the action at weak coupling $U_\mu=\Gamma_\mu$.
We observed no sign of phase transition behaviour  except for low 
values of $b$ and small $N$. Indeed, some of our results extend below 
the transition point between  strong and weak coupling, which was
estimated to be $b=0.3596(2)$ in Ref.~\cite{campostrini}. Results
below this value of $b$ lie then  in a metastable phase (at infinite $N$).
Nonetheless, it seems that as $N$ grows the probability of tunnelling 
decreases considerably and there are no signs of phase flips even down 
to $b=0.35$. Whether the tests of volume independence are performed
in a metastable region or not is of no particular concern, as was the 
case for the  partial reduction studies of Ref.~\cite{NN}. The same 
applies for the other  phases presented in the same reference.  In all 
our simulations we  have monitored the behaviour  of open loops which
are not invariant  under the $Z(\sqrt{N})^4$ symmetry of the reduced 
model and found them to be compatible with zero within errors. As 
mentioned in the introduction, the choice of flux $k$ is obviously 
crucial for this result. Our results, therefore, provide a test that 
the  criteria, presented  in Ref.~\cite{TEK3} to avoid symmetry
breaking, works well.

Finally, a comment about the total statistics and computer resources
employed in our results. The number of different simulation parameters 
involved in our work are in the hundreds, and have been  accumulated 
over several years. Typically, the number of configurations used for 
our large volume
(L=16,32) periodic boundary conditions results is the range 300-600,
and have been generated  with INSAM clusters at Hiroshima University.
For the TEK model we have used  typically  6000 configurations at each 
$b$ value at N=841,  and 2000 configurations at N=1369. The most 
extensive and computer-wise demanding results were obtained with the 
Hitachi SR16000 supercomputers at KEK and YITP.
Some small scale results were obtained with the clusters and servers 
at IFT.

\section{Size and N dependence of small Wilson loops}

\subsection{General considerations about volume independence}
Expectation values of the lattice observables depend on the
conditions of the simulations, namely the values of $b$, $L$, $N$ 
and the twist tensor $n_{\mu \nu}$. Restricting ourselves to the
symmetric twists we may write $O(b,N,L,k)$ for a generic observable $O$.
In the thermodynamic or infinite volume limit, local  observables should 
have a well defined value:
\be
\lim_{L \longrightarrow \infty} O(b,N,L,k) = O(b,N)
\ee
Notice that in that limit the dependence on the boundary conditions
drops out. In general, we expect the infinite volume observable to have 
a well defined large $N$ limit:
\be
\lim_{N \longrightarrow \infty} O(b,N) =O_\infty(b)
\ee
The observable $O_\infty(b)$ is the large N quantity that we are
interested in. 

The question now poses itself about the commutativity of the order of
both limits. What is the result if we take large N limit first, and
then the infinite volume limit? The concept of volume
independence~\cite{EK} gives 
a striking answer to this question. Taken at face value it predicts 
\be
\label{VR}
\lim_{N \longrightarrow \infty} O(b,N,L,k) = O_\infty(b)
\ee
Even before it was formulated this hypothesis, as such, was known 
to be wrong.  Its validity  depends upon the value of $b$ and the
dimensionality of space-time. While presumably correct in two
dimensions it is certainly not true in higher dimensions and
sufficiently large values of $b$. Still the hypothesis is correct for 
$b$ in the strong coupling region $b<b_c(N)$~\cite{EK}\cite{AGAST}. The
point brought into the discussion by Ref.~\cite{TEK3} is about the
dependence of the statement upon the value of $k$. No doubt that the
choice of $k$ has a crucial influence on the results for small $L$. 
The proposal of Ref.~\cite{TEK2} of  choosing $k$ coprime with $\sqrt{N}$ 
avoids the problems found for $k=0$ (periodic boundary conditions) at small
enough  couplings (large $b$).

A modified hypothesis for the periodic boundary condition ($k=0$) case,
known as partial reduction, was introduced by Narayanan and Neuberger~\cite{NN}.
In view of  their results, they conjectured that
volume independence remains true provided $L>L_c(b)$, with $L_c(b)$ a
growing and asymptotically divergent function of $b$. This idea can be 
combined with the use of twisted boundary conditions.  Not surprisingly, 
the effective value of $L_c(b)$, as obtained in finite $N$ studies, seems 
to depend also on $k$~\cite{GANN}. 

However, numerical studies\cite{TIMO}-\cite{TV} disproved the initial hypothesis
that it is enough to take $k$ coprime with $\sqrt{N}=\hat{L}$ to
restore  volume independence, as formulated in Eq.~\ref{VR}, at all values of
the coupling.  In view of this, a modified volume independence
proposal  was put forward in Ref.~\cite{TEK3}:
\be
\label{MVI}
\lim_{\hat{L} \longrightarrow \infty} O(b,N=\hat{L}^2,L,k_{\hat{L}}) = O_\infty(b)
\ee
This implies that the flux has to be scaled as $\hat{L}$ is sent to
infinity. According to the proposal, based on centre symmetry breaking
arguments,  the precise value of $k_{\hat{L}}$ is not relevant provided
$k_{\hat{L}}/\hat{L}$ and
$\bar{k}_{\hat{L}}/\hat{L}$ remain always larger than a certain threshold
($\sim 0.1$). Here $\bar{k}$ is an integer such that $k \bar{k} = 1
\bmod \hat{L}$. 
The restrictions of working with symmetric twists and values of $N$
that are the square of an integer are presumably not indispensable.
They emerge from the simplicity of maintaining as much isotropy as
possible among the different directions of space-time.  

The validity of any volume independence hypothesis has always been
tested indirectly by verifying the $Z^4(N)$ symmetry condition assumed in 
Eguchi and Kawai proof. Although this is valid road-map, nowadays it is
possible to explore directly how a given particular observable behaves
as a function of the different quantities involved. This has the
advantage of informing us of the rate at which the asymptotic limits
are attained, which is a crucial piece of information in rendering 
volume independence a useful tool in numerical studies. In what
follows we will present the results of our analysis taking as
observables the best measured lattice observables: the plaquette and
other small Wilson loops.  One of the good things of the reduction or
volume-independence hypothesis is that it is valid for the lattice
theory, not only the continuum limit. If reduction takes place in the 
scaling region, this will be inherited by the continuum limit.  

\subsection{Behaviour at very weak coupling}
Although  most of our interest has concentrated in the region of $b$
values from which scaling results are obtained, it is interesting to
examine what happens to volume independence at very weak couplings
(large values of $b$). In that region, perturbation theory is a good
approximation to the behaviour of our lattice quantities. Typically one 
gets an expansion of the form 
\be
O(b,N,L,k)= \hat{O}_0(N,L,k) -\sum_{n=1}^\infty \hat{O}_n(N,L,k) \frac{1}{b^n}
\ee
For Wilson loops the leading term is $\hat{O}_0(N,L,k)=1$. 

Let us start by focusing upon the plaquette $\E(b,N,L,k)$. 
The infinite volume coefficients up to order $n=3$ have been computed by
several authors~\cite{dGR}-\cite{Alles}. The corresponding values in 4
dimensions are 
\bea
\hat{\E}_1(N)&=&\frac{1}{8}(1-\frac{1}{N^2})\\
\hat{\E}_2(N)&=&\frac{1}{24}(1-\frac{1}{N^2})(0.12256631-\frac{3}{16 N^2})\\
\hat{\E}_3(N)&=&(1-\frac{1}{N^2})(0.000794223-0.002265487/N^2+0.0023152583/N^4)
\eea
The finite volume corrections for periodic boundary conditions started
to be studied  long time ago. The main difficulty is the presence of
infinitely many gauge inequivalent configurations with zero-action: 
the torons~\cite{gajka}. If one expands around the trivial classical
vacuum $A_\mu=0$ (which dominates the path integral for d=4 and large $N$), 
one must separate the fluctuations into those with 
zero momentum and the rest.
The non-zero momentum contributions to the plaquette were studied in 
Ref.~\cite{CPT}-\cite{HellerKarsch}. They adopt the form:
\be
\hat{\E}^{(p\ne 0)}_1(N,L,k=0)=\frac{1}{8}(1-\frac{1}{N^2})(1-\frac{1}{L^4})
\ee
The corresponding contribution of the zero-momentum degrees of freedom
was computed in Ref.~\cite{Coste}. In 4 dimensions it becomes:
\be
\hat{\E}^{(p=0)}_1(N,L,k=0)=\frac{1}{12}(1-\frac{1}{N^2})\frac{1}{L^4} 
\ee
It is clear that the leading order finite volume corrections do not cancel each 
other  and do not go to zero as $N$ goes to infinity:
\be
\hat{\E}_1(N,L,k=0)=\frac{1}{8}(1-\frac{1}{N^2})(1-\frac{1}{3L^4})
\ee
This kills the volume independence hypothesis  for periodic boundary
conditions at weak coupling, where the  calculation is reliable. 

For twisted  boundary conditions the calculation is very different
since the action has isolated minimum action solutions. In this case 
the result is  given by~\cite{gajka}\cite{Coste}
\be
\hat{\E}_1(N,L,k\ne 0)=\frac{1}{8}(1-\frac{1}{N^2})
\ee
Hence, there is no volume dependence at this order.
This  result, as well as that of periodic boundary conditions, can be 
easily derived since the expectation value of the plaquette can
be obtained  by  differentiating the partition function with respect
to $b$. To this order what matters is just the counting of gaussian
and quartic fluctuations. 

In the same references the volume dependence  of rectangular $R\times T$ 
Wilson loops  to leading order in perturbation theory was also studied.
The infinite volume coefficients up to order $\lambda^2$  were
computed in Ref.~\cite{WWW}.  For loop sizes much smaller than the
lattice size,  the leading finite volume correction from non-zero momentum is  
\be
\delta \hat{W}^{(p\ne 0)}_1(R,T,N,L,k=0) = \frac{1}{8}(1-\frac{1}{N^2})\frac{R^2
T^2}{L^4} +\mathcal{O}(L^{-6}) 
\ee
while the zero momentum one is 
\be
\delta \hat{W}^{(p=0)}_1(R,T,N,L,k=0) = \frac{1}{12}(1-\frac{1}{N^2})\frac{R^2 T^2}{L^4}
\ee
Again they do not cancel each other, leaving a finite volume
correction at all $N$. On the contrary the result for twisted boundary
conditions takes the form
\be
 \hat{W}_1(R,T,N,L,k\ne0) =\hat{W}^{(p\ne 0)}_1(R,T,N=\infty,L\hat{L},k=0) -
 \frac{1}{N^2}\hat{W}^{(p\ne 0)}_1(R,T,N=\infty,L,k=0)
\ee
where $\hat{L}=\sqrt{N}$. The first term is just the result of
periodic boundary conditions at $N=\infty$ and a lattice volume of
$N^2L^4$. Thus as $N$ grows we approach the infinite volume result 
irrespective of the value of $L$. The second term goes like $1/N^2$
in the large $N$ limit, but plays an important role in reducing finite 
volume corrections. This comes because the leading  $L^{-4}$ corrections
cancel each other between the first and second terms, so that  the 
leading finite volume dependence in the sum goes like  $N^{-2}L^{-6}$. 
Notice that, due to the same reason,  adding the $p=0$ contribution to the $k=0$
coefficients of the right-hand side does not alter the result.  
In summary, large N volume independence holds at this order.

In Ref.~\cite{TEK2} the present authors studied the structure of the
Feynman rules for the TEK model  ($L=1$) and argued that volume
independence should hold at all orders of perturbation theory. This
result trivially extends to large $N$, $k\ne0$  models for any $L$. 
The rules are essentially a lattice version of those of non-commutative 
field theory. The finite $N$ corrections  appear in two ways. On one
side the propagators are just the lattice propagators in a box of size 
$L\sqrt{N}$. This relates finite $N$ effects with finite volume
effects. However, in addition the finite $N$ effect comes in as a
non-planar diagram contribution, which is generically exponentially
suppressed with $N$, but might induce also power-like suppressions. 
Furthermore, the coefficients depend on the choice of the flux $k$.
To quantify finite $N$ and finite volume  corrections, including the 
effect of $k$,  a numerical analysis  of the $\mathcal{O}(\lambda^2)$ 
coefficients  would be quite interesting. This study is currently 
under way~\cite{MGPAGAMO}. 

In the absence of explicit  perturbative calculations, we might use numerical 
simulations at small coupling in order to quantify the size of finite
volume corrections at  weak coupling (large values of $b$). In this 
work we used limited resources, by taking $L=1,2,4$ and $N=49$ for both
periodic and  twisted boundary conditions. A more precise study will be
performed to test the higher order analytic calculations.

\begin{figure}     
\includegraphics[width=.9\linewidth]{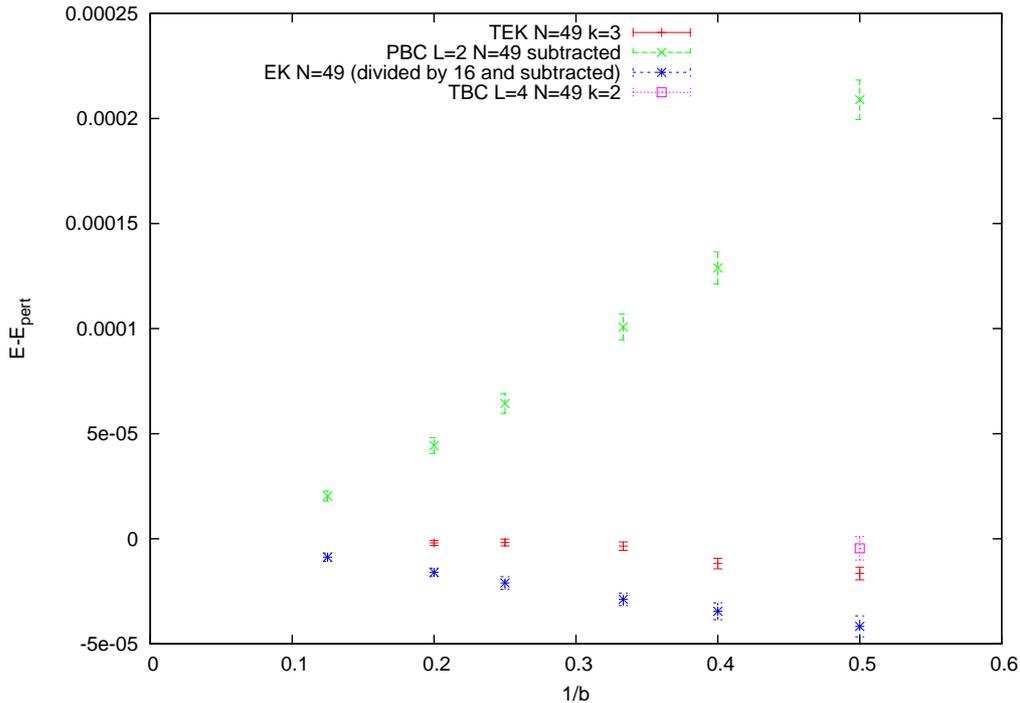}
		   \caption{We plot the measured value of the
		   plaquette $\E$ minus the
		   infinite volume perturbative result up to order
		   $\lambda^3$ for $N=49$.
		   The result for the TEK model (red points) is
		   compatible with zero to
		   within two sigma. For the periodic boundary
		   conditions (L=1 blue and
		   L=2 green) we subtracted the finite size
		   corrections to order $\lambda$
		   and plotted the difference (divided by 16 for
		   $L=1$).  }
		   \label{figpert}
\end{figure}

Hence, we will compare our numerical results with the predictions of
perturbation theory at infinite volume. The results for the plaquette 
expectation value at  $b\ge 2$ and  $k\ne 0$ are 
quite compatible with the perturbative formulas.  For example, the
perturbative
prediction at infinite volume at $b=2$ and $N=49$ is  $0.936152$, while
the measured value for $L=4$ and a symmetric twist with $k=2$ is $0.936147(5)$.
The difference
is compatible with zero within errors, which are  smaller  than the
$1/N^2$ corrections, since the perturbative value at $N=\infty$
gives $0.936123$. For the TEK model ($L=1$) one gets $0.936134(4)$, $0.936140(4)$,
$0.936135(4)$ for $k=1,2,3$ respectively, which is in
agreement with the $L=4$ result and within acceptable $1/b^4$ corrections of the
perturbative value.  We emphasise that the agreement is  sensitive to 
the $\mathcal{O}(\lambda^2)$ and higher perturbative coefficients. 
To see this, we fitted a cubic polynomial in $1/b$ to the plaquette results 
fixing the constant and linear coefficients to the perturbative infinite 
volume values. The best fit (having chi square per degree of freedom less than 1)
gives a value $0.00511(3)$ for the  coefficient in $1/b^2$, while the 
infinite volume coefficient
is $0.005102$. The $1/b^3$ coefficient fits to a value of $0.00091(8)$
which is not far from the infinite volume result $0.000793$. 
 The panorama is summarised in Fig.~\ref{figpert}.
 In it we show the difference between the measured and three-loop
 (order $1/b^3$) infinite volume perturbative  value of the plaquette for 
 the TEK model ($L=1$) $N=49$ and $k=3$. Thus,
within the precision of the numerical data  we  conclude
that the plaquette of the Twisted Eguchi-Kawai model in the perturbative
region is compatible with infinite volume perturbation theory already for $N=49$.

The previous result is in strong contrast with the result for periodic 
boundary conditions. We already found that there are sizeable finite volume 
corrections of order $\lambda=1/b$. However, even if we subtract out this 
contribution
finite size effects are still considerably larger than for TEK. 
In Fig.~\ref{figpert} we show  the case of $L=2$ $k=0$, and
also the $L=1$ $k=0$, which has been divided by 16 to fit it into the same
plot.

\begin{figure}
\includegraphics[width=.9\linewidth]{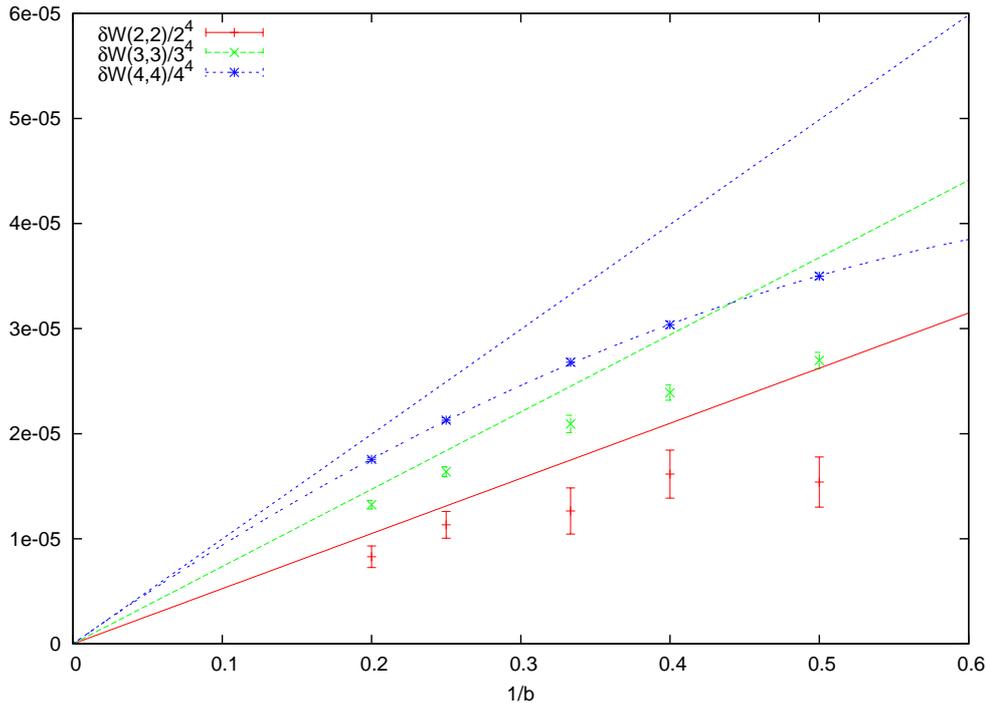}
\caption{For the TEK model at $N=49$ and $k=3$ we plot $\delta W(R,R)$,
the difference between the measured
expectation values of Wilson loops and the next-to-leading perturbative
formula at infinite volume. The result is divided by $R^4$ and compared
with the analytic prediction at leading order (straight line in the
same colour) and a second-degree polynomial fit for $R=4$.}
\label{fig2pert}
\end{figure}

It is also very interesting to look at what happens for larger loops. 
In contrast with the case of the plaquette, the result at lowest order 
has a  non-vanishing finite size correction which, as we saw earlier, is of
order $1/N^2$. The question is whether there is numerical evidence that 
higher orders violate this rule and produce large corrections. Again
we have looked at the TEK model at $N=49$ and various $b\ge 2$ values. 
We measured the mean values of the  Wilson loops and subtracted 
from it the perturbative formula for the infinite volume Wilson loop, 
up and including  order $\lambda^2$. Hence, the difference $\delta
W(R,T)$ measures the leading corrections to volume independence.  
Asymptotically  this must be dominated by the term linear in $1/b$
which can be computed analytically, and which for very large $N$ and 
$R,T \ll \sqrt{N}$ is 
given by 
\be
-\frac{\lambda}{N^2} (\hat{W}_1(R,T,N=\infty) - \frac{R^2 T^2}{8}) +
\mathcal{O}(\frac{1}{N^3})
\ee
Thus, the correction for $R=T$  grows with the fourth power of
$(R/\sqrt{N})$. As expected $\sqrt{N}$ acts as the effective 
size of the box. Thus,  in Fig.\ref{fig2pert} we plot our result for
$\delta W(R,R)/R^4$ at $N=49$  with twisted boundary conditions and $k=3$. 
The straight lines are the leading order perturbative contribution. 
Since, $R=T=3,4$ are not much smaller than $\hat{L}=7$ the slopes 
for the various $R$ change by a factor of 2. The actual data lie
always below the corresponding straight line, implying that higher
powers of $1/b$ produce a dependence of the opposite sign. Indeed, 
one can easily describe the data points by simply adding a $1/b^2$
correction with a coefficient  determined by a fit. The corresponding
line is also plotted as the blue curve for $R=T=4$.

\subsection{N-dependence of Plaquette at   $b=0.36$}
In what follows we move to the other edge of the weak coupling phase
of the large N model. We choose a particular value of the coupling
$b=0.36$, right above the strong to weak phase transition point.
At this point we carried a large number of simulations to determine  
the expectation value of the plaquette for various values of $N$, $L$ 
and the flux $k$. As mentioned earlier, we expect  that when $L$ is
large enough, the result  should be independent of the boundary conditions
and given by $\E(0.36,N)$. We estimate that at $L=16$ one gets a good measure
of  this quantity within the errors of order $0.00001$. In particular,
for $N=8$, $L=16$ and $k=0$ (periodic boundary conditions) one 
gets $\E(0.36,8,16,k=0)=0.572022(19)$, which is compatible with the value 
obtained at $N=8$, $L=32$ and $k=0$ given by $\E(0.36,8,32,k=0)=0.572019(5)\sim
\E(0.36,8)$.

To test the volume independence hypothesis,
we performed a series of simulations in a
lattice of size $L^4=16^4$ with periodic boundary conditions, for $b=0.36$
and all values of $N$ in the range $N\in[8,16]$. The mean plaquette 
value results  are  displayed in Fig.~\ref{plaquette036a} as a
function of $1/N^2$.  The dependence on $N$ is clearly visible, 
being much larger than the individual errors, which are too small to be 
seen in the plot. A good fit is obtained by a second  degree
polynomial in $1/N^2$. The $\chi^2$ per degree of freedom is 0.248,
for 9 points and 3 parameters. The coefficient of the linear term is
$0.960(3)$. The constant coefficient  in the fit provides the extrapolated
value of the plaquette at $N=\infty$, and its value is 
$\E_\infty(0.36)=0.558012(12)$. 
To test the $L$ dependence we repeated the analysis in an $8^4$ and
 $4^4$ periodic box. The results of the $8^4$ box are numerically
 close to those of $16^4$, but the difference is much larger than the 
 errors and clearly seen in the 
plot. For $N=8$ the difference is $3.41(70)\, 10^{-4}$ and for $N=16$ it is
$2.42(50)\, 10^{-4}$. A similar three parameter fit describes the data 
very well and gives an extrapolation of $0.558114(67)$.  Indeed, even
fixing the extrapolated value to be equal to the $L=16$ one, we  get a
fit with a chi square per degree of freedom which is smaller than 1. 
Thus, our results are consistent with the basic claim that the volume
dependence drops to zero in the large $N$ limit. 

In contrast we also
show the results of the $L=4$ periodic lattice. For this smaller
volume we have explored values of $N=8,10,12,14,16,25,49,81$. A good
polynomial fit can be obtained which undoubtedly extrapolates to a very
different value in the large $N$ limit. Thus, volume independence
fails in this case. This is precisely what we expected on the basis of
the results of Narayanan and Neuberger~\cite{NN}. At $b=0.36$ volume
independence should only work for $L\ge L_c(0.36)\sim 8$. 

\begin{subfigures}
\begin{figure}
\centering
%\begin{subfigure}{0.45\textwidth}
%  \centering
    \includegraphics[width=\linewidth]{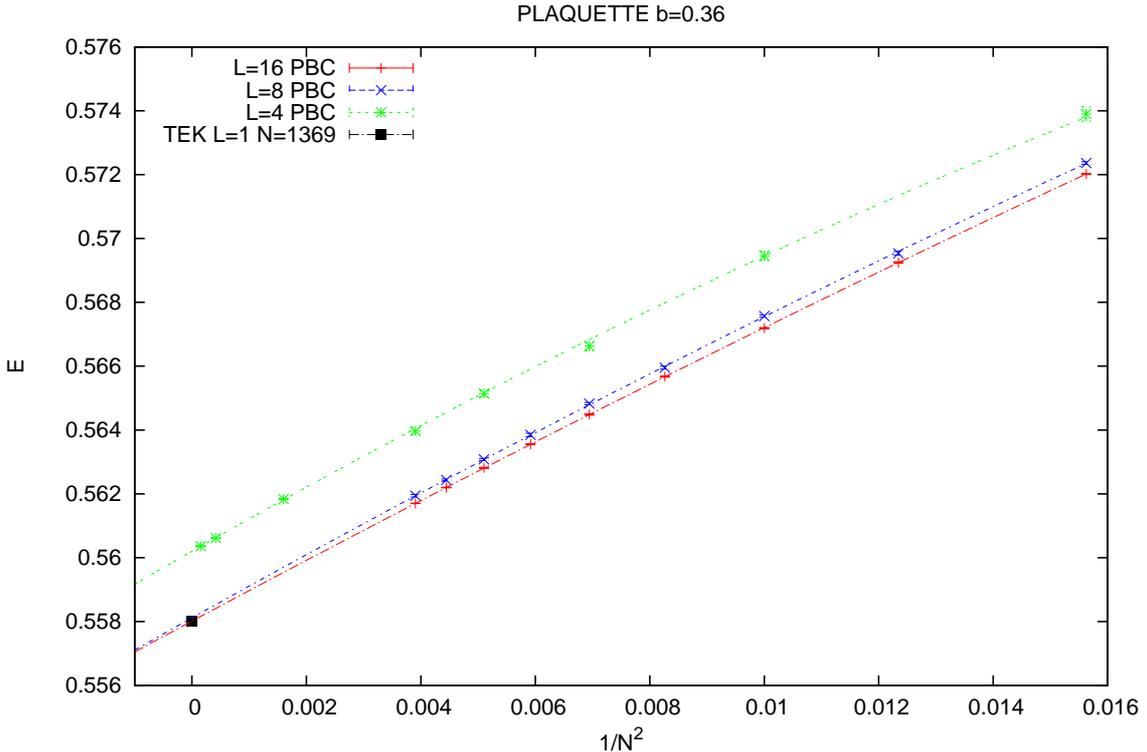}
          \caption{The expectation value of the plaquette $\E$ for
	        $b=0.36$ is plotted as a function of $1/N^2$. We used periodic
		      boundary conditions on a box of size $L^4$ with $L=4,8,16$.
		            The black point is the result of the TEK
			    model for $k=11$ and
			          $N=1369$. }
\label{plaquette036a}
\end{figure}
					  %       \end{subfigure}%
					  %       \begin{subfigure}{0.45\textwidth}
\begin{figure}
%         \centering
\includegraphics[width=\linewidth]{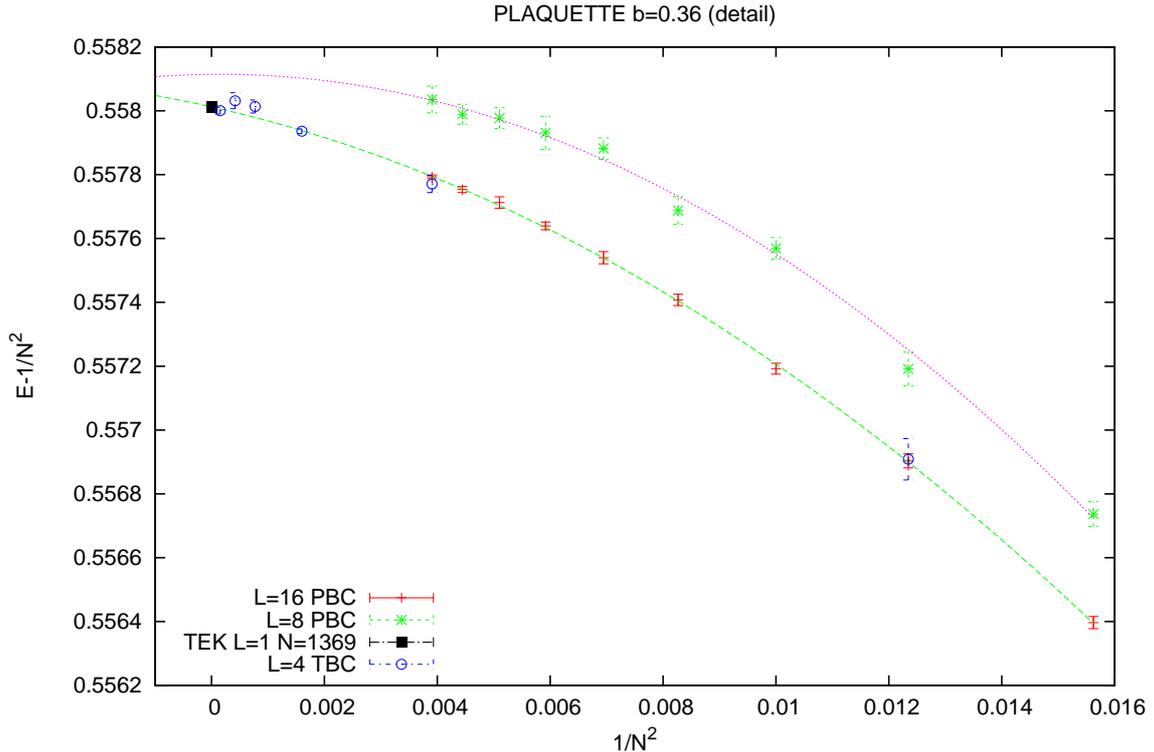}
\caption{We display
the same $L=16,8$ data as in Fig.~\ref{plaquette036a} but subtracting $1/N^2$. In
	               addition we also plot the TEK data point at
		       $N=1369$ and
           various values obtained at $L=4$
	   and symmetric twist.}
	   \label{plaquette036b}
	   \end{figure}
						     %               \end{subfigure}
						     %
						     % \caption{Figure 1}
						     %               \label{fig:test}
						     %               \end{figure}
\end{subfigures}

The important role of boundary conditions, as claimed in
Refs.~\cite{TEK1}-\cite{TEK2}, is certainly supported by our data. The
black point in Fig.~\ref{plaquette036a} is our result with the TEK model, 
namely a one point box ($L=1$) with symmetric twisted boundary
conditions, flux $k=11$ and group rank $N=1369$. From a single simulation we 
get $\E_\infty(0.36)=0.558019(11)$, which is in perfect agreement with the
extrapolated result. Obviously, it is to be expected that the same
applies if one uses partially reduced situations with $L>1$, provided
one uses appropriate boundary conditions. Indeed, we studied the $L=2$
and $L=4$ cases with symmetric twisted boundary conditions and
$N=9,16,25,36,49,81$ (with $k$ values of $1,1,2,1,2,2$ respectively). 
Here the N dependence of the result is quite 
obvious, but follows the same curve as for the $L=16$ periodic boundary
conditions one. We take this curve to  give the $N$ dependence at
infinite volume $\E(0.36,N)$. The leading correction, as mentioned
earlier, is a $1/N^2$ term with a coefficient close to one. We use
this fact to display the data in a different fashion, which allows to
get a better grasp of the precision involved. Plotting $\E-1/N^2$ the
errors of the $L=16,8$ data are clearly seen in the
plot (Fig.~\ref{plaquette036b}).
Notice the small scale on the y axis, showing the high precision of
the data. The curves shown are the afore-mentioned fits to the 
periodic boundary condition data in the range $N\in[8,16]$. Once again
we see that the curve  goes through the TEK point at
$N=1369$, whose errors are of the size of the point. The results of
$L=4$ with twisted boundary conditions are also shown and are
consistent with the same curve. This shows that  $L=4$ is close enough
to infinite volume when twisted boundary conditions are used even for 
$N$ as small as 9. Our $L=2$ data with tbc do not disrupt the picture.
Indeed, a simultaneous fit to the 
$L=2,4$ twisted and $L=16$ periodic box has chi square per degree of
freedom smaller than 1.

The previous data provide the most precise test of volume independence
to date. Errors in the plaquette are of order $10^{-5}$, and to that
level the Twisted Eguchi-Kawai model matches the extrapolated result. 
At this level of precision the $1/N^2$ correction  is still sizeable
up to $N\sim300$. We emphasise that this agreement is putting to test the
validity of reduction in the non-perturbative regime. The
value of $b=0.36$ lies in the region in which scaling is best observed 
and corresponds to a lattice spacing of $0.1$ in $\Lambda_{\rm MS}$
units. Thus, it is hard
to argue against the fact that nonperturbative phenomena contribute
corrections to  the  plaquette higher than $10^{-5}$. 

\begin{figure}[ht]
\centering
%\begin{minipage}{.45\textwidth}
%  \centering
%\hskip -1cm
    \includegraphics[width=\linewidth]{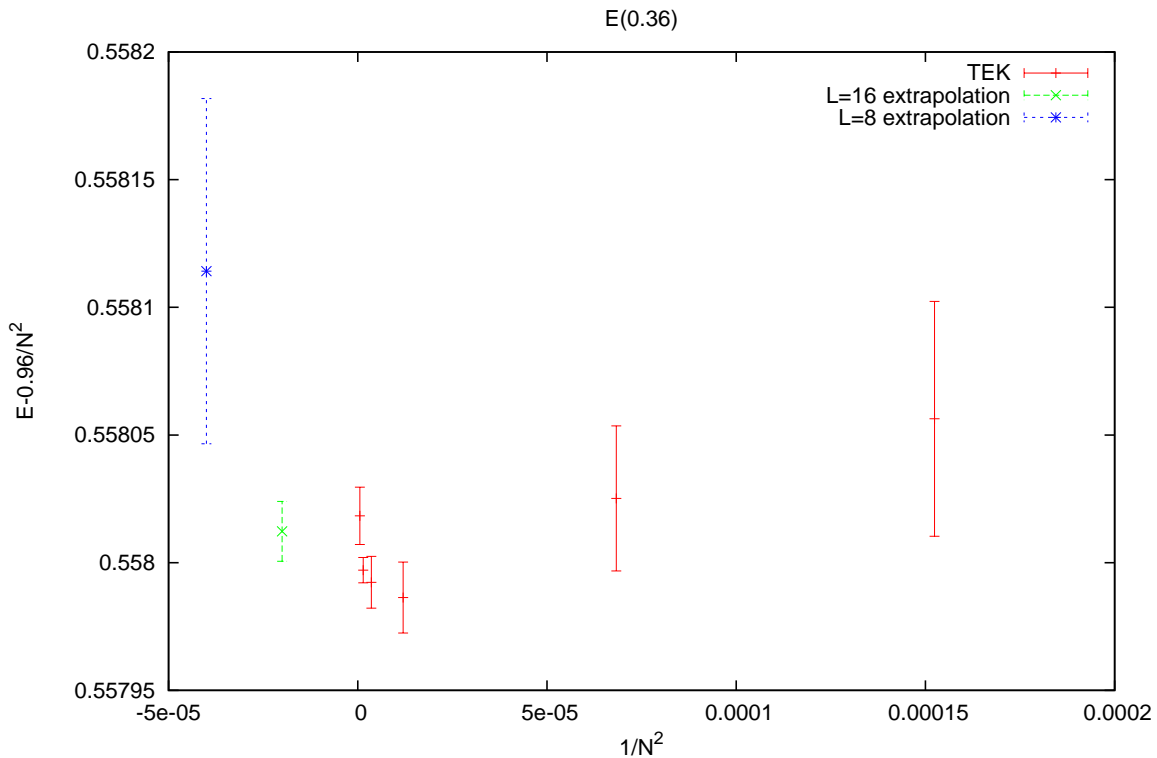}
    %\hskip -1cm
    \caption{We plot the plaquette $\E$ minus $0.96/N^2$
          as a function of $1/N^2$ for the TEK model. The green point is
	        the infinite $N$ extrapolated value from $L=16$ data. The
		      blue point on the left, the corresponding result
		      for $L=8$. }
		              \label{figextrap}
			      \end{figure}

In any case, as argued in the introduction, our goal goes beyond the
yes/no answer to volume reduction. We want to investigate the size and 
nature of the errors affecting physical quantities at large but finite
$N$. For the case of the plaquette we have seen that there is a
leading $1/N^2$ correction which is pretty much volume independent
provided one uses sufficient large sizes and/or appropriate boundary
conditions. For example, at $L=16$ and $N=16$ the plaquette is 
$E(0.36,16)=0.561700(6)$, which is still relatively far from the
infinite $N$ value of $0.558$. Using twisted boundary conditions
($k=1$) and  $L=4$ the result is $E(0.36,16)=0.561712(20)$. For 
$L=2$ one has $E(0.36,16)=0.561748(84)$. Thus the three results are
perfectly consistent with each other. Unfortunately  the $L=1$ twisted
result for this $N$ cannot be obtained due to the vicinity to the 
strong coupling phase.  The plaquette shows frequent jumps to a low 
value around $0.4$. As mentioned in the previous section these flips 
are suppressed at larger values of $N$. 

We have seen that the $N$-dependence  of  the plaquette
expectation value for large lattice sizes  matches with  the result of small 
$L=2,4$ sizes and symmetric twists. In principle, 
there is no known reason why the $1/N^2$  corrections  should not 
depend on $L$ and the value of $k$. For the case of the TEK model
($L=1$) it is important to know how big these corrections are, in order 
to optimise the values of $N$ used to reproduce infinite volume
results within a given precision. 
To investigate this matter, we studied the TEK 
model for $N=81,121,289,529,841,1369$ corresponding to
$\hat{L}=9,11,17,23,29,37$. The values of the flux adopted were
$k=2,3,5,7,9,11$ respectively. These choices are dictated by our criteria
to avoid symmetry breaking. The resulting values of the plaquette are
collected in Fig.~\ref{figextrap}. We plot $E(0.36,N)-0.96/N^2$
according to the value of the leading correction obtained from the 
$L=16$ periodic boundary condition data. This correction only affects
sizably the results of $N=81$ and $121$. What we see is that all the
results are consistent with each other within errors. That shows the 
approximate universality of the $1/N^2$ correction. On the negative side
of the x-axis at arbitrary locations we placed the prediction for 
the infinite volume, infinite $N$, plaquette $E_\infty(0.36)$ obtained by 
extrapolation of the $L=16$ and $L=8$ periodic boundary conditions
data. The results are consistent, and give a best estimate 
$E_\infty(0.36)=0.558002(5)$. The different values and errors are also
displayed in Table~\ref{table1}. 

\begin{figure}[ht]

	    \includegraphics[width=\linewidth]{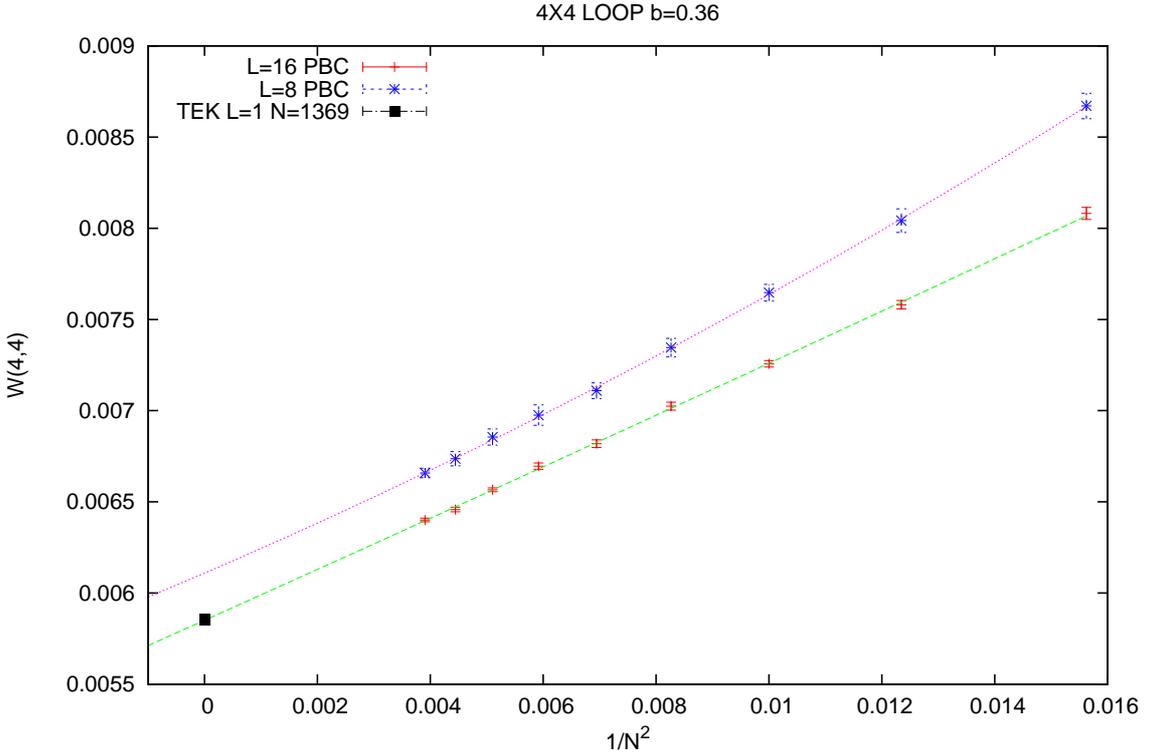}
	      \caption{This is similar to Fig.~\ref{plaquette036a}
	      but for the $4 \times 4$ Wilson loop. \vspace*{2cm}}
	        \label{figwloop}
		
%		\end{minipage}
		\end{figure}

We should emphasise that the choice of $k$ is important. If we approach 
those values for which symmetry breaking is observed, significant 
departures are observed. For example, taking $N=81$ and $k=1$, the
plaquette becomes $0.5554(1)$. Although this is less than 1\% away from 
the result at $k=2$, it is many standard deviations away given the 
precision of our data. 

\subsection{Results for other small loops and for other values of $b$}
Having seen that reduction works at the level of the tiny statistical 
errors for the plaquette, we should also explore other quantities. In
principle, extended quantities should be more sensitive to finite
volume corrections than the plaquette. Hence, we explored also the 
results for square loops of linear size $R$ (The $R=1$ case is just
the plaquette). The relative errors grow considerably with $R$.
Typically, errors remain more or less constant, but the value of the
observable drops considerably; at $R=4$ by a factor of a hundred. 
Since, we want to test the observables with the minimum amount of
manipulation we restrict our analysis to $R\le 4$. 

The methodology is similar to the one for the plaquette. The $L=16$ 
results are fitted to a quadratic polynomial in $1/N^2$ allowing
extrapolation to large N. The coefficient of the $1/N^2$ is $1.04(1)$,
$0.44(2)$ and $0.14(1)$ for $R=2,3,4$ respectively. For the larger loops
the quartic coefficient is compatible with zero.
The large $N$ extrapolated values are collected in Table~\ref{table1} together
with the plaquette values. The fits are good and match with the result of 
the TEK model. As an example, the situation for the largest loop $W(4,4)$ is 
shown in Fig.~\ref{figwloop}. Notice that, as a consequence of the small 
expectation value of the loop,  the $1/N^2$ correction gives a quite 
significant contribution in relative terms (of order 10\% at $N=16$).  
The results for $L=8$ periodic boundary conditions  give compatible 
extrapolations for all loops except for $R=4$ as seen clearly 
from  the figure.

\begin{table}
\begin{tabular}{||l||c|c|c|c||}\hline \hline
MODEL & Plaquette $\E_\infty$ & $W_\infty(2,2)$ & $W_\infty(3,3)$ &
$W_\infty(4,4)$ \\ \hline \hline
L=16 PBC & 0.558012(12) & 0.155294(37) & 0.032533(52) & 0.005851(26)  \\ \hline
L=8 PBC & 0.558114(67) & 0.155569(112) & 0.032543(97) & 0.006110(23) \\ \hline
TEK N=1369 &  0.558019(11) & 0.155256(23) & 0.032489(14) &
0.005861(13) \\ \hline
TEK N=841  & 0.557998(5) & 0.155209(10) & 0.032430(7) & 0.005835(5) \\ \hline
TEK N=529 & 0.557996(10) & 0.155215(15) & 0.032418(13) & 0.005853(10)\\ \hline
TEK N=289 & 0.557998(14) & 0.155235(21) & 0.032399(16) & 0.005841(12) \\ \hline
L=4 TBC & 0.558022(9) & 0.155257(21) & 0.032488(24) & 0.00586(40) \\ \hline
L=2 TBC & 0.55797(1) & 0.155186(71) & 0.03239(13) & \\ \hline \hline
\end{tabular}
\caption{Estimates of the infinite volume large N 
plaquette $\E_\infty(0.36)$ and small square loops $W_\infty(R,R)$ obtained by
 extrapolation of $L=16$ and $L=8$ periodic boundary conditions (PBC) data.
 The same observables are given for the TEK ($L=1$) model and other
 partially reduced symmetric twist results with $L=2,4$. All data correspond
 to $b=0.36$. }
 \label{table1}
\end{table}

\begin{figure}
 \includegraphics[width=.9\linewidth]{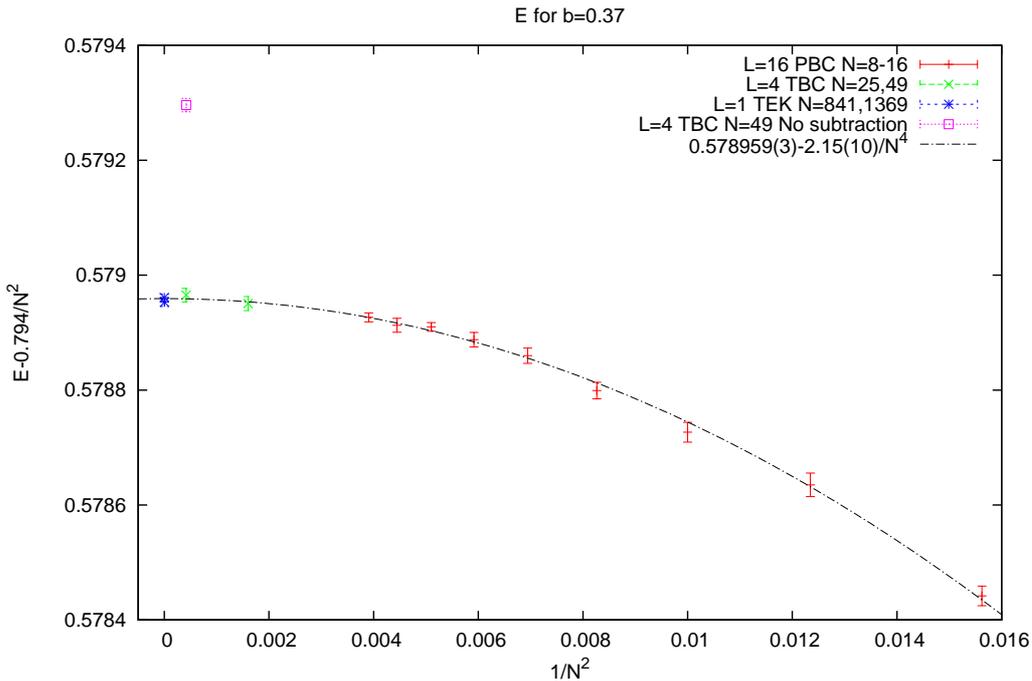}
   \caption{The plaquette $\E$ for $b=0.37$  after subtracting a
     $1/N^2$ correction. Red points come from $L=16^4$ lattice with
       periodic boundary conditions. The two overlapping blue points are
         TEK simulations. The green ones are from a partial twisted
	 reduction
	   with $L=4$ and $N=25,49$. The latter point is also plotted without
	     the $1/N^2$ subtraction. }
	       \label{plaquette037}
	         \end{figure}

Hence, the evidence that at $b=0.36$ reduction is at work with errors
of order $10^{-5}$ extends to Wilson loops of size up to $4\times 4$.
A less extensive study has been carried at other values of $b$. 
For example, at $b=0.37$ and $L=16$ we also measured the same
observables in the range $N\in[8,16]$. Again the data is beautifully
described by a second degree polynomial in $1/N^2$. The linear
coefficient is now $0.789(5)$. This gives an estimate of
$\E_\infty(0.37)=0.578978(17)$, which compares quite well with the direct
measurements of the TEK model at $N=1369$ and $N=841$ which were 
$\E_\infty(0.37)=0.578961(6)$ and $\E_\infty(0.37)=0.578954(7)$ respectively. 
The situation is depicted in Fig.~\ref{plaquette037} in which we also
included the results of partially reduced twisted simulation with
$L=4$, $k=2$ 
and $N=25,49$ which fit nicely into the same curve. Here we are not
only testing the $L$ independence but also the approximate
universality of the leading $1/N^2$ correction. Even for $N=49$ this
correction is non-negligible. To show this, we also displayed the $L=4$
$N=49$ point without subtraction, which gives the square which stands
outside the curve. To the level of the errors it is clear that the
$1/N^2$ correction is rather large. 

\begin{figure}
 \includegraphics[width=.9\linewidth]{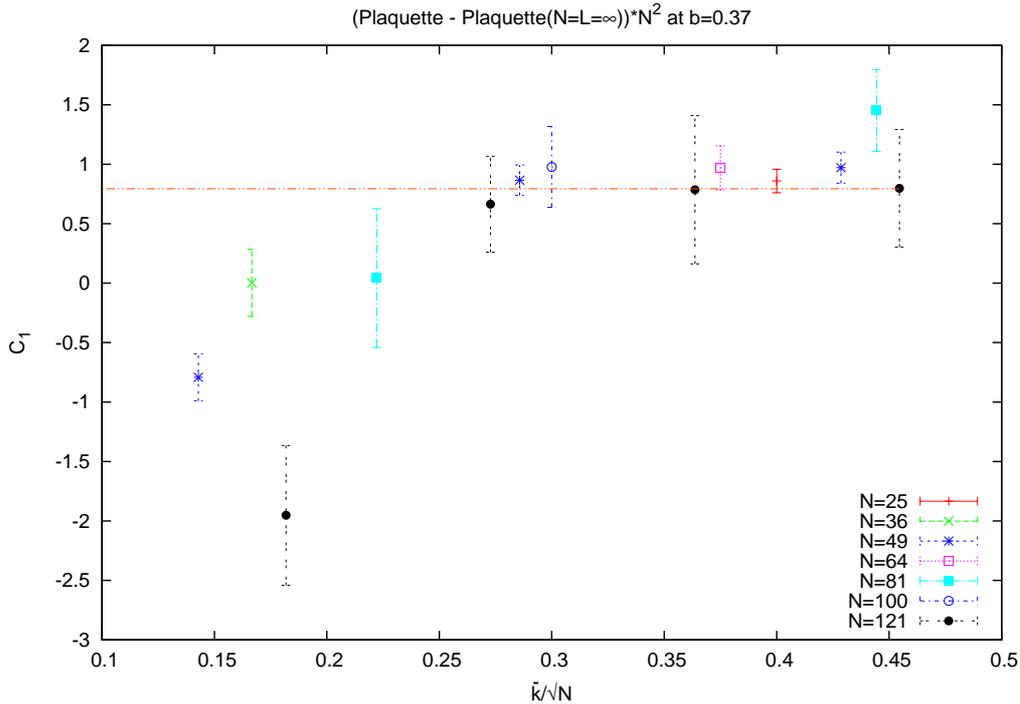}
   \caption{We plot the  coefficient $C_1$ defined in Eq.~\ref{C1EQ}
   as a function of $\bar{k}/\hat{L}$ for various simulations of the
   TEK model at $b=0.37$ and various values of $N$ and $k$.  }
   	       \label{kdepfig}
	         \end{figure}

We also analysed the characteristic $1/N^2$ dependence for the TEK model
($L=1$). Obviously for that one has to simulate small values of $N$,
since for $N >300$ the correction is smaller than the statistical errors.
Furthermore, it is to be expected that the coefficient depends on the
flux $k$. Thus, we simulated the model for various $(k,N=\hat{L}^2)$
combinations. In all cases we computed the coefficient $C_1$ given by 
\begin{equation}
\label{C1EQ}
C_1=N^2(\E(0.37,N,L=1,k)-\E_\infty(0.37))
\end{equation}
The results are plotted in Fig.~\ref{kdepfig} as a function of 
$\bar{k}/\hat{L}$  (we recall that $\bar{k}k= 1 \bmod \hat{L}$). 
Within the allowed region ($>0.1$) 
the coefficient lies between $-2$ and $2$, implying that there are no 
unexpectedly large $1/N^2$ corrections in any case. Furthermore, 
all values having $\frac{\bar{k}}{\hat{L}}>0.25$ are consistent with
the infinite volume  $1/N^2$  coefficient  marked as a horizontal line
in the plot.

For bigger loops the $1/N^2$ is rather a finite volume correction,
which should grow as the loop size approaches the effective lattice 
size $\hat{L}$. Nonetheless, the equivalent $C_1$ coefficient remains 
within reasonable limits. For example, for $4\times 4 $ loops the 
coefficient remains in the band $[-6,10]$ for $\bar{k}/\hat{L}>0.1$.

The most important qualitative difference  of the $b=0.37$ data with 
respect to the $b=0.36$ one is that the $L=8$
periodic boundary condition data does not seem to extrapolate to the
same value. The result is $0.57935(7)$ which is not terribly far away,
but inconsistent within errors. Indeed, this was to be expected on the
basis of the Narayanan and Neuberger results, since at $b=0.37$ the
$L=8$ size lies outside the unbroken symmetry region.

The previous procedure was repeated for various other values of $b$.
Two estimates of the large $N$ infinite volume Wilson loop 
expectation values  can be  given. One is the large $N$
extrapolation of the results on an $16^4$ periodic box. The other
comes from simulating the TEK model at various large values of $N$.
The TEK results are consistent among themselves and compatible with
the ones coming from large $N$ extrapolation. Furthermore, they tend
to have smaller errors. Table~\ref{table2} summarises the results.  

\begin{table}
\begin{tabular}{||l||c|c|c|c||}\hline \hline
MODEL & Plaq. $\E_\infty$ & $W_\infty(2,2)$ & $W_\infty(3,3)$ & $W_\infty(4,4)$ \\ \hline \hline
Average TEK  b=0.350	 & 0.529672( 1)	 & 0.123504( 4)	 & 0.019199( 15)	 & 0.002319( 11)	 \\ \hline
Extrapolated b=0.350	 & 0.529846( 40)	 & 0.123689( 66)	 & 0.019182( 35)	 & 0.002340( 31)	 \\ \hline \hline
Average TEK  b=0.355	 & 0.545336( 11)	 & 0.140781( 14)	 & 0.026071( 37)	 & 0.003999( 12)	 \\ \hline
Extrapolated b=0.355	 & 0.545417( 63)	 & 0.140926( 64)	 & 0.026103( 42)	 & 0.003937( 83)	 \\ \hline \hline
Average TEK  b=0.365	 & 0.569018( 4)	 & 0.168113( 5)	 & 0.038554( 8)	 & 0.007813( 6)	 \\ \hline
Extrapolated b=0.365	 & 0.569021( 41)	 & 0.168096( 28)	 & 0.038592( 59)	 & 0.007796( 27)	 \\ \hline \hline
Average TEK  b=0.370	 & 0.578959( 5)	 & 0.180040( 11)	 & 0.044547( 6)	 & 0.009926( 3)	 \\ \hline
Extrapolated b=0.370	 & 0.578978( 17)	 & 0.180129( 32)	 & 0.044573( 46)	 & 0.009966( 45)	 \\ \hline \hline
\end{tabular}
\caption{Two estimates of the large $N$ infinite volume expectation
values of square loops for various $b$. One estimate is obtained by  
extrapolation to infinite  $N$ of results on a $L=16$ box with periodic
boundary conditions. The other is the direct measurement with
the TEK model and various $N$.}
\label{table2} 
\end{table}

\subsection{Global Fit to small loop expectation values}
In the previous subsections we have analysed the validity of the
reduction idea by comparing the results of infinite volume gauge
theories with those obtained for the TEK model. Our first tests were
obtained in a region dominated by  perturbation theory, and then 
we moved to the opposite edge of the weak coupling region. It is
tempting to use all the results to obtain a global description at all
values of $b$.  For practical purposes it is also interesting to have a 
formula that allows to interpolate the various values of $\E_\infty(b)$ 
that have been collected in the previous tables. This can allow us to make
use of all the available data on the TEK model and SU(N) gauge theories 
generated at various values of $b$ in the course of previous
investigations~\cite{TEK4}-\cite{TEK5}. 

If we analyse all plaquette expectation values obtained for the TEK model, 
it is remarkable that all our results for $b\le 0.385$ fall neatly into a 
straight  line in the variable $1/b_I$, where 
$b_I=b*\E_\infty(b)$. The improved coupling $b_I$ was introduced by
Parisi~\cite{Parisi} and used extensively in Ref.~\cite{ATT}. 
The function $0.89661(32)-0.006803(7)/b_I$ approaches the numerical
values at the level of $0.0001$. Nonetheless, the fit cannot be
considered  good, since the errors in the plaquette values are one order 
of magnitude smaller.

This led us to look for another parameterization, which could
not only fit the measured values but also  match with the perturbative
expression up to three loops  studied earlier. It was not hard to
find a solution as a ratio of two polynomials of third degree in
$1/b_I$. The constant term of both polynomials is 1. Of the remaining
six parameters, three are fixed by the perturbative formula. Fitting
the remaining three parameters to the TEK values we get a fit with 
chi-square per degree of freedom equal to 0.22. Given the nice
properties of the formula and the high precision of the data, this
seems quite remarkable. 

We tried to extend the previous formula to the determination of
$\E(b,N)$. As mentioned earlier, in addition to the $L=16$ data used for 
the extrapolations presented in previous subsections, 
we do have a good amount of data coming from our 
determination of the large $N$ string tension~\cite{TEK4}-\cite{TEK5}. 
Altogether, we have 88 points for $L=16$ and $L=32$ with periodic
boundary conditions, various values of $b$ and $N$ ranging from 3 to 16.
A good fit is obtained to all of them by replacing the 3 parameters of
the previous parameterization by a second degree polynomial in $1/N^2$. 
The constant term is fixed to the parameters of the large $N$ fit.
Altogether, there are 6 new parameters, although some are clearly
redundant and one can obtain a good fit  having a chi square per degree
of freedom of $\sim 0.7$ with only 4 parameters. Furthermore, $1/N^4$ terms
are only necessary if we include the results at $N=3$ and $N=4$. 

Summarising, we may write
\be
\label{bestfit}
\bar{\E}(b,N)=\frac{N(x,y)}{D(x,y)}
\ee
where $x=1/b_I$ and $y=1/N^2$. 
The numerator $N(x,y)$ is a polynomial
of the form 
\be
N(x,y)=1+\sum_{n=1}^3 \sum_{m=0}^{\min(2,n)} a_{nm} x^n y^m
\ee
The denominator has exactly the same form with different coefficients 
$a'_{nm}$ which are completely determined by the $a_{nm}$ and the
coefficients of the perturbative expansion of the plaquette to order 
$1/b^3$ (we leave the details to the reader). Notice that
Eq.~\ref{bestfit} gives $\bar{\E}(b,N)$ in an implicit way, since
$b_I=b\bar{\E}(b,N)$. Nevertheless, the formula allows one to solve 
analytically for $\bar{\E}(b,N)$ as a function on $b$ with a formula
that involves radicals.

\begin{figure}%[b!]
\begin{center}
%\resizebox{4in}{!}
{
% GNUPLOT: LaTeX picture with Postscript
\begingroup
  \makeatletter
  \providecommand\color[2][]{%
    \GenericError{(gnuplot) \space\space\space\@spaces}{%
      Package color not loaded in conjunction with
      terminal option `colourtext'%
    }{See the gnuplot documentation for explanation.%
    }{Either use 'blacktext' in gnuplot or load the package
      color.sty in LaTeX.}%
    \renewcommand\color[2][]{}%
  }%
  \providecommand\includegraphics[2][]{%
    \GenericError{(gnuplot) \space\space\space\@spaces}{%
      Package graphicx or graphics not loaded%
    }{See the gnuplot documentation for explanation.%
    }{The gnuplot epslatex terminal needs graphicx.sty or graphics.sty.}%
    \renewcommand\includegraphics[2][]{}%
  }%
  \providecommand\rotatebox[2]{#2}%
  \@ifundefined{ifGPcolor}{%
    \newif\ifGPcolor
    \GPcolortrue
  }{}%
  \@ifundefined{ifGPblacktext}{%
    \newif\ifGPblacktext
    \GPblacktexttrue
  }{}%
  % define a \g@addto@macro without @ in the name:
  \let\gplgaddtomacro\g@addto@macro
  % define empty templates for all commands taking text:
  \gdef\gplbacktext{}%
  \gdef\gplfronttext{}%
  \makeatother
  \ifGPblacktext
    % no textcolor at all
    \def\colorrgb#1{}%
    \def\colorgray#1{}%
  \else
    % gray or color?
    \ifGPcolor
      \def\colorrgb#1{\color[rgb]{#1}}%
      \def\colorgray#1{\color[gray]{#1}}%
      \expandafter\def\csname LTw\endcsname{\color{white}}%
      \expandafter\def\csname LTb\endcsname{\color{black}}%
      \expandafter\def\csname LTa\endcsname{\color{black}}%
      \expandafter\def\csname LT0\endcsname{\color[rgb]{1,0,0}}%
      \expandafter\def\csname LT1\endcsname{\color[rgb]{0,1,0}}%
      \expandafter\def\csname LT2\endcsname{\color[rgb]{0,0,1}}%
      \expandafter\def\csname LT3\endcsname{\color[rgb]{1,0,1}}%
      \expandafter\def\csname LT4\endcsname{\color[rgb]{0,1,1}}%
      \expandafter\def\csname LT5\endcsname{\color[rgb]{1,1,0}}%
      \expandafter\def\csname LT6\endcsname{\color[rgb]{0,0,0}}%
      \expandafter\def\csname LT7\endcsname{\color[rgb]{1,0.3,0}}%
      \expandafter\def\csname LT8\endcsname{\color[rgb]{0.5,0.5,0.5}}%
    \else
      % gray
      \def\colorrgb#1{\color{black}}%
      \def\colorgray#1{\color[gray]{#1}}%
      \expandafter\def\csname LTw\endcsname{\color{white}}%
      \expandafter\def\csname LTb\endcsname{\color{black}}%
      \expandafter\def\csname LTa\endcsname{\color{black}}%
      \expandafter\def\csname LT0\endcsname{\color{black}}%
      \expandafter\def\csname LT1\endcsname{\color{black}}%
      \expandafter\def\csname LT2\endcsname{\color{black}}%
      \expandafter\def\csname LT3\endcsname{\color{black}}%
      \expandafter\def\csname LT4\endcsname{\color{black}}%
      \expandafter\def\csname LT5\endcsname{\color{black}}%
      \expandafter\def\csname LT6\endcsname{\color{black}}%
      \expandafter\def\csname LT7\endcsname{\color{black}}%
      \expandafter\def\csname LT8\endcsname{\color{black}}%
    \fi
  \fi
  \setlength{\unitlength}{0.0500bp}%
  \begin{picture}(7200.00,5040.00)%
    \gplgaddtomacro\gplbacktext{%
      \csname LTb\endcsname%
      \put(814,704){\makebox(0,0)[r]{\strut{}-5}}%
      \put(814,1111){\makebox(0,0)[r]{\strut{} 0}}%
      \put(814,1518){\makebox(0,0)[r]{\strut{} 5}}%
      \put(814,1925){\makebox(0,0)[r]{\strut{} 10}}%
      \put(814,2332){\makebox(0,0)[r]{\strut{} 15}}%
      \put(814,2740){\makebox(0,0)[r]{\strut{} 20}}%
      \put(814,3147){\makebox(0,0)[r]{\strut{} 25}}%
      \put(814,3554){\makebox(0,0)[r]{\strut{} 30}}%
      \put(814,3961){\makebox(0,0)[r]{\strut{} 35}}%
      \put(814,4368){\makebox(0,0)[r]{\strut{} 40}}%
      \put(814,4775){\makebox(0,0)[r]{\strut{} 45}}%
      \put(946,484){\makebox(0,0){\strut{} 3.8}}%
      \put(1635,484){\makebox(0,0){\strut{} 4}}%
      \put(2324,484){\makebox(0,0){\strut{} 4.2}}%
      \put(3013,484){\makebox(0,0){\strut{} 4.4}}%
      \put(3702,484){\makebox(0,0){\strut{} 4.6}}%
      \put(4391,484){\makebox(0,0){\strut{} 4.8}}%
      \put(5080,484){\makebox(0,0){\strut{} 5}}%
      \put(5769,484){\makebox(0,0){\strut{} 5.2}}%
      \put(6458,484){\makebox(0,0){\strut{} 5.4}}%
      \put(176,2739){\rotatebox{-270}{\makebox(0,0){\strut{}($E-\bar{E})*10^5$}}}%
      \put(3874,154){\makebox(0,0){\strut{}$\frac{1}{b_I}$}}%
    }%
    \gplgaddtomacro\gplfronttext{%
      \csname LTb\endcsname%
      \put(1606,4602){\makebox(0,0)[r]{\strut{}TEK}}%
      \csname LTb\endcsname%
      \put(1606,4382){\makebox(0,0)[r]{\strut{}N=16}}%
      \csname LTb\endcsname%
      \put(1606,4162){\makebox(0,0)[r]{\strut{}N=8}}%
      \csname LTb\endcsname%
      \put(1606,3942){\makebox(0,0)[r]{\strut{}N=4}}%
      \csname LTb\endcsname%
      \put(1606,3722){\makebox(0,0)[r]{\strut{}N=3}}%
    }%
    \gplbacktext
    \put(0,0){\includegraphics{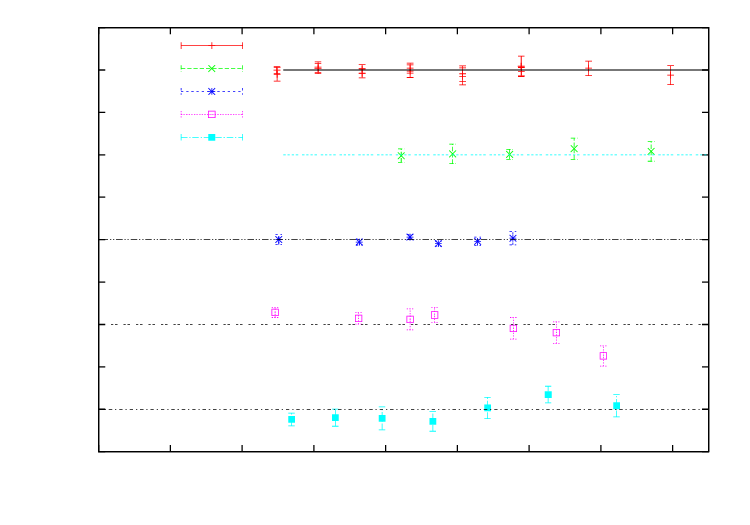}}%
    \gplfronttext
  \end{picture}%
\endgroup

}
%\captionsetup{font=small, labelfont=bf, labelsep=period}
%\caption[Small caption]{LARGE CAPTION}
\caption{Comparison of our data points with the best fit function 
Eq.~\ref{bestfit} for a selection of values of $N$. Data of 
different $N$ are  shifted by multiples of 10  for display purposes.
From top to bottom TEK, $N=16$, $N=8$, $N=4$ and $N=3$.}
\label{figFITN}
\end{center}
\end{figure}

In Fig.~\ref{figFITN} we show the difference of our plaquette measurements 
and the function Eq.~\ref{bestfit} for certain  values of $N$. The 
different values of $N$ are shifted by multiples of 10, for
a better presentation.  The best fit parameters are given by 
 $a_{10} =-0.0644631$, $a_{20} =-0.0162287$, $a_{30} =0.0005949$,
 $a_{11} =0.1312897$, $a_{21} =-0.0143626$, $a_{31} =-0.0002531$,
$a_{22} =-0.1871392$ and $a_{32} =0.0042532$.

The conclusion of this result is that the whole data set is perfectly
consistent. All values can be fitted with a simple function whose 
large $N$ behaviour has been determined entirely in terms of the
TEK measured values. The finite $N$ values are dominated by the $1/N^2$ 
correction which describes the data from a relatively small value of
$N$ on.

\begin{figure}
  \centering
    \includegraphics[width=\linewidth]{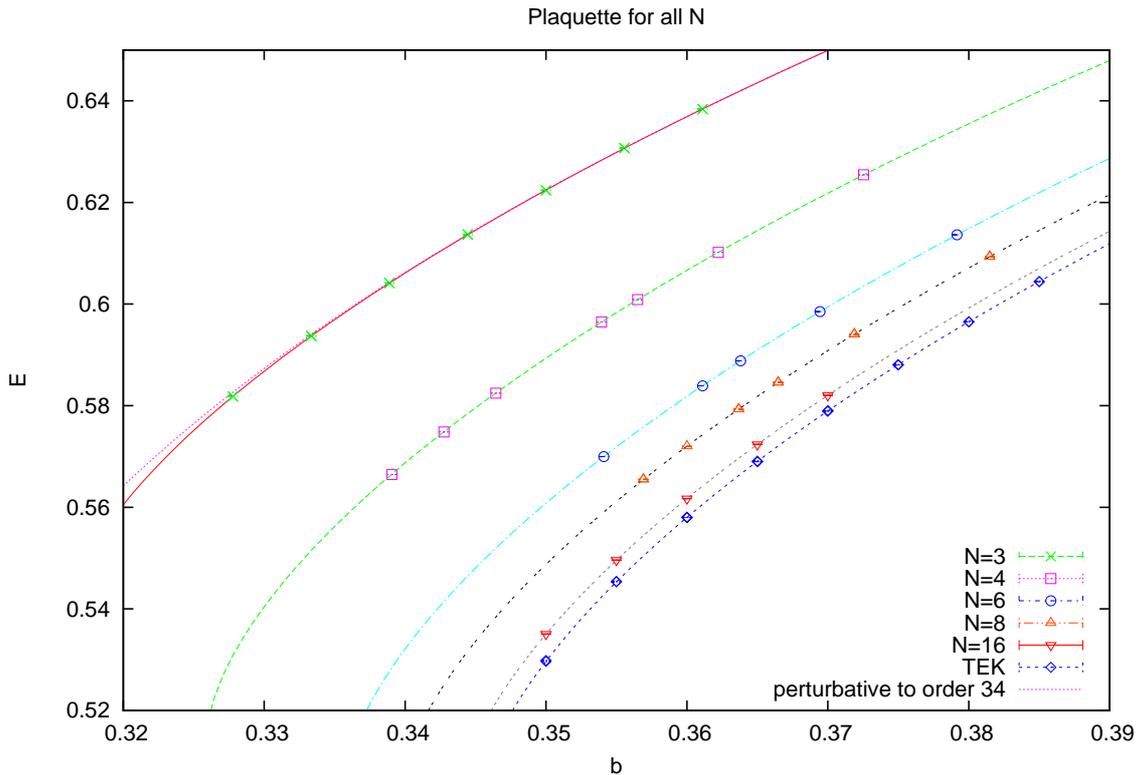}
        \captionof{figure}{We display the plaquette expectation value
	$\E$
	    as a function of $b$ for a sample of our data points. The finite
	        $N$ results were obtained in simulations of boxes with
		linear size
		    $L=16,32$. The $N=\infty$ points are given by the
		    TEK model. The
		        curves are our best fit. }
			    \label{plaquette}
\end{figure}

Furthermore, having this formula allows one to estimate the
plaquette expectation value at intermediate values. For example, we
could compare our formula with the measurements of other authors at
large volumes and different  couplings. We always found agreement to
the level of $\pm 0.00002$. Nonetheless, one must be cautious about
using our formula for extrapolation  outside of the  fitting range.
In principle,  for large $b$ this is less worrisome, since the
function incorporates the  perturbative behaviour up to order $1/b^3$.
To give a bird's eye view of our formula and the data we show the
plaquette value $\E$ as a function of $b$ for several values of $N$ 
in Fig.~\ref{plaquette}. To the scale of the figure our best fit curves 
go right through the centres. For comparison we added also the curve 
reconstructed from the 34 perturbative coefficients obtained recently 
for the SU(3) case~\cite{bali}. At this scale this curve also fits the 
points and you only start to see visual deviations of both curves for
low b values.

\begin{figure}
    \centering
    \includegraphics[width=\linewidth]{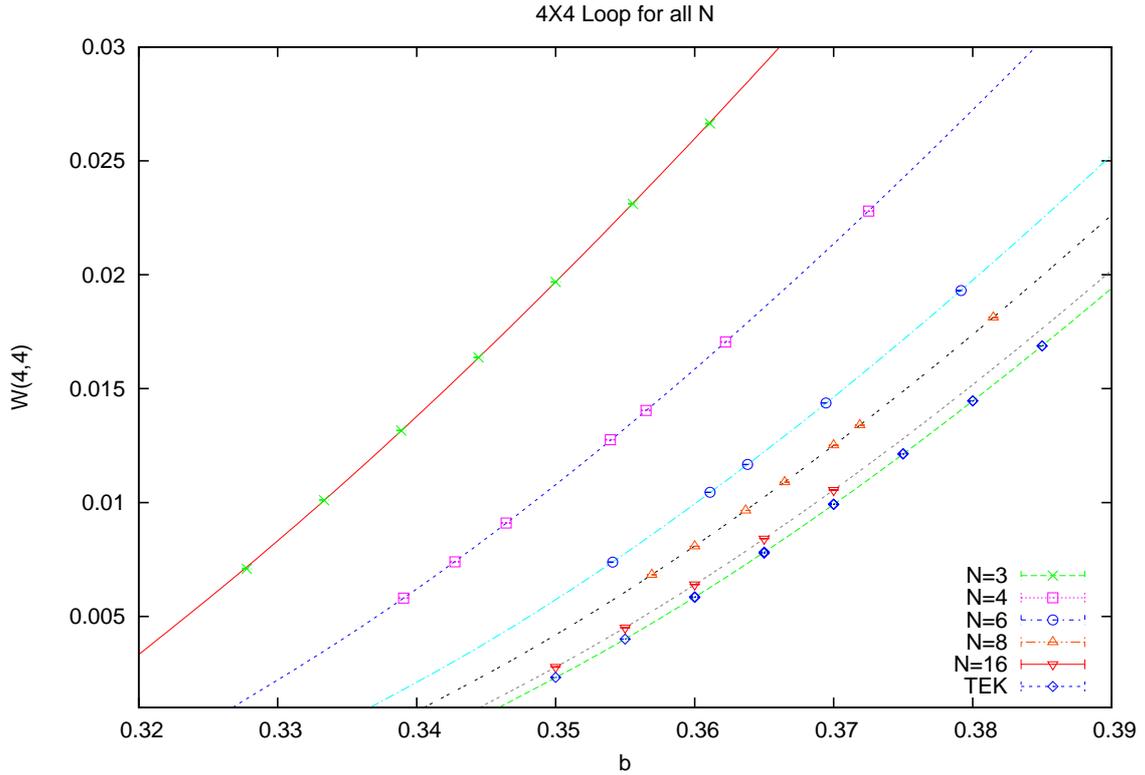}
    \captionof{figure}{Similar to the previous figure but for the
    $4\times 4$ Wilson loop. \vspace*{2.7cm}}
    \label{wloop4}
\end{figure}

The same procedure can be extended  to  square loops up to $4\times 4$.
There are a bunch of small differences in our procedure and also 
in the quality of the results. We considered more reasonable to fit
the logarithm of the Wilson loops. This time, however, we lack the 
$\mathcal{O}(1/b^3)$ perturbative coefficients for all $N$, and we used
only the results  of Ref.~\cite{WWW}. In order to deal with this 
lack of information without enlarging the number of parameters we
used a (3,2) Pad\'e approximant with coefficients that are polynomials 
in $1/N^2$. The lowest order 3  parameters are determined in terms of
the TEK data only and give good fits. The higher order terms in $1/N^2$ 
are determined from the SU(N) data with poorer fits having chi-squares 
of 2 times the number of degrees of freedom (points-parameters). Given
the small errors there are still  tiny differences between the
measured values and the fits. In Fig.~\ref{wloop4} we give as an
example  the bird's eye view of the $4\times 4$ Wilson loop, which at
this scale shows no visible deviation. 

As explained earlier, the interest of the fit is that we have been
able to make use of all the data 
obtained during the last few years by our group both for the TEK
reduced model as for ordinary SU(N) lattice field theory  at various
values of $N$. The main conclusion is that everything fits  into a 
picture in which  the TEK
results  match nicely as the large N extrapolation of ordinary gauge
theory at infinite volume for the  wide range of couplings explored.

%\end{document}

\section{Conclusions}
In this paper we have presented the results of an extensive analysis
of the expectation values of small Wilson loops 
for SU(N)  Yang-Mills theory on the lattice with Wilson action. Our 
analysis allows to test the dependence of these observables on the
rank of the group $N$, the size of the box $L$ and the boundary
conditions. Our results provide direct evidence that, under certain 
conditions, the large $N$ results are independent of the lattice size
$L$. For this to happen the choice of boundary conditions is crucial. 
Twisted boundary conditions are most effective in reducing the volume 
dependence at large $N$, as advocated in Ref.~\cite{TEK1}-\cite{TEK2}. 
Our data are consistent with the claim of Ref.~\cite{TEK3}, that 
even the one point model (TEK) captures the infinite volume large $N$ results
if the large $N$ limit is taken with a suitable choice of the fluxes. 
Hence, our conclusion is that an appropriate version of volume 
independence holds. Although this is not a mathematical proof, the 
observables studied can be determined with  a high accuracy. 
Hence, the validity of the statement has been tested down to the 
level of the statistical errors of our data, which are of order $10^{-5}$.

We have analysed the region of large values of $b=1/\lambda$ where 
perturbation theory is a good approximation, but most of our results 
were obtained in the interval $b\in[0.35,0.385]$ typically used to
extract   continuum limit results. In any case, there is no evidence 
for any pathological behaviour at any value of $b$. Indeed, we were 
able to find a parameterization of the infinite volume 
plaquette and small square loop expectation values which encodes the 
known perturbative behaviour at weak coupling and fits all our measured 
values. Furthermore, the large $N$ limit of this parameterization
matches nicely with the results of the TEK reduced model. A large
amount of data obtained over the last few years has gone into this
analysis.

As stressed in the introduction the goal was not only to check the 
volume independence hypothesis but also to estimate its corrections. 
At infinite volume the Wilson loops that we studied approach 
the large $N$ limit with a leading correction that goes as $1/N^2$
with a coefficient (depending on the value of $b$) of order 1. 
The correction is not small at the scale of the errors ($\sim 10^{-5}$) 
and  only becomes comparable for values of $N\sim 200-300$.
Thus, performing simulations at various $N$ and
extrapolating can hardly be avoided. Furthermore, for the plaquette, 
the typical $1/N^2$  corrections of the reduced model  have similar size
to those of SU(N) lattice gauge theory. Hence, a single simulation of the
reduced model  gets the correct result with orders of  magnitude less 
degrees of  freedom. In all our previous statements care has to be taken on the 
choice made for the flux integer $k$. However, our results do not 
show problematic dependencies provided one stays within the safe region 
proposed in Ref.~\cite{TEK3}.

For larger loops one must take into account the connection between 
the $\sqrt{N}$ and an effective lattice size which follows from 
perturbation theory. Hence, deviations are  expected if the loop 
sizes become close to $\sqrt{N}$. Here, we have shown that provided 
one remains away from this situation, things look pretty much like 
for the plaquette. 

 In this work we intended to  verify the commutativity of the 
large $N$ and large volume limits for typical and precise lattice
quantities. Our test does not rely on centre symmetry and Eguchi-Kawai 
proof. Things are nevertheless consistent since our order parameters
for the symmetry are compatible with zero in all our simulations. 
If reduction holds for the lattice model, this property should be 
inherited by the corresponding continuum limits. Indeed, this 
track was followed earlier~\cite{TEK5} when we attacked the 
determination of the continuum string tension at large $N$.
Although our results were strikingly consistent with volume
independence, they employed sophisticated methods of noise reduction 
for large loops as well as extrapolation to the  continuum limit. 
This paid a price from the point of view of simplicity, as well as
considerably enlarging the errors. 
Hence, the present work can be seen as a necessary complement.

\section*{Acknowledgments}
We thank Margarita Garc\'{\i}a P\'erez for useful conversations and a 
critical reading of the manuscript.

A.G-A acknowledges financial support from the grants FPA2012-31686
and FPA2012-31880, the MINECO Centro de Excelencia Severo Ochoa Program SEV-
2012-0249, the Comunidad Aut\'onoma de Madrid HEPHACOS S2009/ESP-1473, and
the EU PITN-GA-2009-238353 (STRONGnet). He participates in the Consolider-
Ingenio 2010 CPAN (CSD2007-00042). 
M. O. is supported by the Japanese MEXT grant No 26400249. 

Calculations have been done on Hitachi SR16000 supercomputer both at High Energy 
Accelerator Research Organization(KEK) and YITP in Kyoto University.  Work at 
KEK is supported by the Large Scale Simulation Program No.14/15-03. 
Calculations have also been done on the INSAM clusters
at Hiroshima University and the HPC-clusters at IFT.

We want to pay tribute to the memory of Misha Polikarpov and Pierre
van Baal who were  early pioneers working on this field and 
from which we benefitted through conversations over the years. 

\bibliographystyle{JHEP}

%\end{document}

%\begin{thebibliography}{}

%\end{thebibliography}

\end{document}